\documentclass{icldt}

\usepackage{amsmath}
\usepackage{amssymb}
\usepackage{MnSymbol}
\usepackage{wasysym}
\usepackage{epigraph}

\title{Stress-Energy in the Conical Vacuum and its Implications for Topology Change}
\author{Eric Jones}
\date{September 18, 2015}
\department{Physics}
\supervisor{Prof. H. F. Dowker}
\dedication{To My Parents}

\begin{document}

%%\fontfamily{cmtt}\selectfont

\maketitle

\begin{abstract}

%%Abstract Text Goes Here
This dissertation presents a semiclassical analysis of conical topology change in $1+1$ spacetime dimensions wherein, to lowest order, the ambient spacetime is classical and fixed while the scalar field coupled to it is quantized. The vacuum expectation value of the scalar field stress-energy tensor is calculated via two different approaches. The first of these involves the explicit determination of the so called Sorkin-Johnston state on the cone and an original regularization scheme, while the latter employs the conformal vacuum and the more conventional point-splitting renormalization. It is found that conical topology change seems not to suffer from the same pathologies that trousers-type topology change does. This provides tentative agreement with conjectures due to Sorkin and Borde, which attempt to classify topology changing spacetimes with respect to their Morse critical points and in particular, that the cone and yarmulke in $1+1$ dimensions lack critical points of unit Morse index.

\end{abstract}

\begin{acknowledgements}

I would foremost like to thank Fay Dowker, whose depth of knowledge, ready support, and contagious sense of wonder about the universe have been elemental to the completion of this project. For guidance and many fruitful discussions, much gratitude is due Ian Jubb. The work of Michel Buck and initial conversations with him facilitated an easy apprehension of the background on this subject. And it will be evident that much of the foundation for this analysis has been laid by Rafael Sorkin. To these two and all of the other authors whose work I have drawn on, I extend my thanks.

Immeasurable love and thanks to my Dad, Darrell, who got this whole ball rolling when he first lent me his copy of ``A Brief History of Time,'' my Mom, Leah, who has always inspired me to find beauty in the world, my brother, Trevor, who has been a stalwart intellectual companion over the years, and the rest of my friends and family for their encouragement and support. I thank Matthias Beuerle, Iacopo Russo, Marlene Weber, Nick Duch, Colin Price, and Manuel Costa Campos for their friendship, and Meera Patel, John Ronayne, Azin Khan, and Eduardo Garc\'ia for lighting up the computational suite with their colorful dispositions.

\end{acknowledgements}

\makededication

\tableofcontents
%\listoftables
%\listoffigures

\chapter{Introduction}
% \epigraph{To proceed during the daytime, because appearances and all outer and inner phenomena, which arise like diverse reflections in a mirror, are nothing more than radiant manifestations of emptiness which have no intrinsic self-nature, there is nothing one should really consider to exist.}{Ch\"ogyal Namkhai Norbu, Dream Yoga and the Practice of Natural Light}
\section{The Zero}
The existence, and corresponding conception or perception, of nothingness is a notion that has beleaguered natural philosophers back into antiquity. Even the semantic contradiction inherent in using ``existence'' and ``nothingness'' in the same breath highlights the crux of the issue as posed by Parmenides of Elea in the fifth century BCE and paraphrased by Bertrand Russel \cite{ nla.cat-vn3538300}: \textit{``You say there is the void; therefore the void is not nothing; therefore it is not the void.''} Already, another term has been introduced to try and clarify things. Indeed, atomists such as Democritus postulated that the void must exist to allow atoms to move through it and to occupy the volumes between them, but this position was retrograded by Aristotle and finally Newton who delineated between an ``eternal space'' upon which physical matter played out its dynamics and the philosophical concept of the void, or absolute nothingness.

Furthermore, we now know through the paradigm of general relativity, that this eternal space is anything but. Rather, it is dynamical and ever changing, rendering its candidacy for playing the role of void substantially more suspect. And finally, when one considers the ground state of quantum electrodynamics, with its roiling sea of virtual particles, continuously materializing and dematerializing, the idea that physical nothingness could be identified with metaphysical nothingness becomes outright untenable.

However, we might still run some thought experiment wherein all of the physical fields of the Standard Model are banished and we are left with pure spacetime and perhaps a scalar field coupled to it acting as a sort of indicator field. Is there some modest sense in which we can recover the concept of void as it relates to this system? The physical situation to which we refer of course is the \textit{vacuum state}, and the theoretical framework we will use is that of semiclassical gravity. As it turns out, for a general classical spacetime, the answer to this question is still not straightforward. And when one tries to incorporate some proposed features of quantum gravitational spacetimes into the analysis, things become even further nettled with uncertainty.

\section{Gravitational Topology Change}

\subsection{Geons and the Quantum Foam}

The great triumph of Einstein's relativity was the discovery not only that spacetime is a geometric object, but that it is also a dynamical one. However, for the purpose of being able to describe a complete classical physics, matter still has to be introduced by hand through parameters such as mass and charge. In 1955, John Archibald Wheeler proposed the concept of the geon, or ``gravitational electromagnetic entity'' \cite{PhysRev.97.511}. The hope was that these self-cohering geometric structures could be used in lieu of the comparatively ad hoc ``bodies'' of classical physics in order to provide: ``gravitation without gravitation, electromagnetism without electromagnetism, charge without charge, and mass without mass'' \cite{misner1957classical}. In order to achieve charge without charge, for example, electromagnetic field lines are confined to pass through a wormhole into two disparate regions of spacetime, giving the macroscopic appearance of source charges. It soon became apparent though, that if a non-trivial topological structure such as this were to exist in a spacetime, physicists might also need a topological dynamics in order to complement the geometrodynamics.

In order to see why this might be, consider another concept due to Wheeler, that of the quantum foam. Classical spacetime is described as a smooth manifold endowed with a Lorentzian metric, $(M,g_{\mu \nu})$. Once the matter content of the physical system under consideration is specified, the dynamics are determined through the Einstein or Einstein-Maxwell field equations. So long as we should like to retain a Lorentzian metric everywhere, the spatial topology of the spacetime must be fixed in the absence of closed timelike curves \cite{gerochctc}. However, suppose we concentrate our attention towards a very small patch of spacetime over a very short period of time at around the Planck length, $\ell_p \sim 1.6 \times 10^{-33}$ cm and time, $t_p \sim 5.4 \times 10^{-44}$ s. On this scale, quantum gravitational effects are thought to dominate the dynamics of spacetime. In particular, if we interpret the energy-time uncertainty relation, $\Delta E \Delta t \sim \hbar$ as allowing the temporary existence of virtual particles with arbitrarily high energies over commensurately short timescales, then at the Planck scale, existence of these virtual particles should cause violent fluctuations to the stress-energy tensor, which in turn ought to result in correspondingly large fluctuations in the metric. Wheeler argued that such agitation to the continuum manifold could cause topology change at the quantum level \cite{1957AnPhy...2..604W}. This is colloquially referred to as the quantum foam.

While seemingly disallowed classically, topology change \textit{can} be admitted by relaxing the requirement that $M$ have an \textit{everywhere} Lorentzian metric, and allowing the metric to degenerate at a finite set of points \cite{Sorkin:1989ea}. Moreover, topological processes could occur as virtual processes at the quantum level, entering into the gravitational amplitude as a phase in the path integral $Z= \int_{geometries} e^{\frac{i}{\hbar}S_{geometries}}$ \cite{pathintegral}. Just as the non-relativistic and relativistic path integrals allow particle paths and field configurations which do not obey Hamilton's principle, to contribute phases to the amplitude, it is not unthinkable that a similar situation might be in view for classically forbidden topological contributions in gravity.

Finally, we can also ask how one could go about formulating a consistent theory of quantum geons, and in fact, this was one of the initial motivations for Wheeler's introduction of the classical geon. The modern term for the relevant physical entity is the \textit{topological geon}, and it has been shown that in the absence of spatial topology change, topological geons as fundamental particles suffer from violations of the spin-statistics theorem \cite{topologicalgeons}.

\subsection{String Theory}

Perhaps the contemporary theory of quantum gravity which makes most ready use of topology change is string theory. This can be seen from the very outset within the context of bosonic string theory. In the simplest scenario, let $(M,\delta_{\mu \nu})$ be 26 dimensional Euclidean space after a Wick rotation from $(M,\eta_{\mu \nu})$, and $(\Sigma_g,\gamma_{\alpha \beta})$ a Riemann surface of genus $g$ embedded therein with coordinates, $X^{\mu}(\xi^0,\xi^1)$. The string action is given by \cite{opac-b1120046}:

\begin{equation}
S[X,\gamma,g]:=\frac{1}{4 \pi \alpha'} \int d^2\xi \sqrt{\gamma}\gamma^{\alpha \beta} \partial_{\alpha} X^{\mu} \partial_{\beta} X^{\nu} \delta_{\mu \nu} + \lambda \chi_g.
\end{equation}

The first term is the Polyakov action with Regge slope $\alpha'$, while $\chi_g=\frac{1}{4\pi} \int d^2 \xi \sqrt{\gamma} \mathcal{R}=2-2g$ is the Euler characteristic, which is a topological invariant on $\Sigma_g$, and $\mathcal{R}$ is the scalar curvature on $\Sigma_g$. The naive vacuum amplitude of the theory would read:

\begin{equation}
Z=\sum_{g=0}^{\infty} \int \mathcal{D}X \mathcal{D}\gamma e^{-S[X,\gamma,g]}.
\end{equation}
It turns out that the correct partition function is:
\begin{equation} \label{eq:polyvac}
Z=\sum_{g=0}^{\infty} \int_{\mathcal{E} \times M_g} \frac{\mathcal{D}X \mathcal{D}\gamma}{V(Diff \ast Weyl)} e^{-S[X,\gamma,g]},
\end{equation}
owing to the fact that the functional integration measure needs to be normalized to take into account the gauge invariance of $S[X,\gamma,g]$ under both diffeomorphisms, $f:\Sigma_g \rightarrow \Sigma_g$ and Weyl rescalings, $\gamma_{\alpha \beta} \rightarrow e^\phi \gamma_{\alpha \beta}$. If $Diff(\Sigma_g) \ast Weyl(\Sigma_g)$ is the semi-direct product space of the groups of these transformations on $\Sigma_g$, then $V(Diff \ast Weyl)$ is the infinite volume of this space. So, each term in \eqref{eq:polyvac} integrates over all possible metric configurations on, and embeddings of, a given Riemann surface modulo geometries equivalent under the gauge group.

Crucially, the vacuum amplitude also sums over Riemann surfaces of every genus with $e^{-2\lambda g}$ playing the role of perturbative expansion parameter\footnote{Schematically, the expansion looks like: (jelly doughnut) $+$ (glazed doughnut) $+$ (double-handed doughnut)$+ \cdot \cdot \, \cdot$}. When one moves towards incorporating external legs, one simply stitches ingoing and outgoing strings to these topologies. According to bosonic string theory then, topology change in the 26 dimensional vacuum is a fairly pedestrian occurrence. From the point of view of world sheet conformal field theory, the free Polyakov action describes the field theory of 26 scalar fields on a 2 dimensional spacetime. Upon quantization, one of these degrees of freedom always remains as the scalar dilaton under $SO(24)$ \cite{polchinski2001string}.

\subsection{Low Dimensions}

The fundamental closed string vertex then looks rather similar to 26 uncoupled massless scalar fields propagating on the so called ``trouser spacetime,'' which allows an $S^1 \rightarrow S^1 \times S^1$ spatial topology change. The trousers fall into the category of topological cobordisms\footnote{Mathematically, given two $n$ dimensional manifolds M and N, with distinct topologies, a cobordism is an $n+1$ dimensional compact manifold W such that $\partial W=M \sqcup N$ \cite{Borde:1994tx}.} mentioned above whose metric can be chosen to degenerate at a single point. In this instance, the degeneracy occurs at the ``crotch'' as viewed with the trousers hanging upside down and time moving upwards. Various analyses of the trouser spacetime have showed that calculation of the stress-energy tensor of a massless scalar field yields an unrenormalizable divergence at the crotch, which propagates into the future lightcones therefrom \cite{manoguetrousers} \cite{michelthesis}.

While this may seem to work against topology change on the quantum level as well, it is interesting to note two points. First, in the case of the trousers the metric is ``causally discontinuous'' \cite{Dowker:1997hj}. That is, the forward lightcone of every point on the trousers is unique, except at the crotch where it bifurcates in two with one branch leading up each leg. There is a conjecture related to this fact due to Sorkin, which attributes such infinite energy bursts to causal discontinuities in a spacetime as opposed to just irregularities in the metric. The second observation is that the trousers cobordism, as a $1+1$ dimensional submanifold of a $3+1$ dimensional product spacetime with $2$ flat codimensions, is argued to be exponentially suppressed in any sum-over-histories formulation of quantum gravity by an extra factor of $e^{-\frac{A}{4G}}$ over its typical phase suppression due to its distance from stationarity in the Einstein-Hilbert action. $G$ is Newton's constant and $A$ is the area subtended by the flat $2$ codimensional submanifold \cite{Louko:1995jw}.

There is another type of cobordism in two dimensions however, the yarmulke. Along with its time-reverse: a structure we will refer to as the cone, and the trousers, these three cobordisms constitute all of the irreducible $1+1$ dimensional cobordisms. In contrast to the trousers, the cone not only is causally continuous, but according to Louko and Sorkin, its amplitude in the SOH is expected to be exponentially enhanced by $e^{\frac{A}{4G}}$ in the same $3+1$ product manifold scenario described in the preceding paragraph \cite{Louko:1995jw}. The aim of this thesis then will be to determine the behavior of a massless scalar field on the Lorentzian cone and to contextualize the results with respect to Sorkin's conjecture.

%%cite michel's dissertation here
%%points from rafaels paper
%%any 2d cobordism can be constructed from yarmulke and trousers
%%Trousers suppressed in path integral while yarmulke enhanced

%connect to trousers and semiclasscal gravity!!
%would be interesting if scalar qft on trousers had something to say about the dilaton on fundamental string vertex!!

%%Begin Chapter Two
\chapter{Preliminaries}
% \epigraph{O the grey dull day! It seemed a limbo of painless patient consciousness through which souls of mathematicians might wander, projecting long slender fabrics from plane to plane of ever rarer and paler twighlight, radiating swift eddies to the last verges of a universe ever vaster, farther and more impalpable.}{James Joyce, A Portrait of the Artist as a Young Man}

Here we present the mathematical tools we require in order to go about evaluating conical-type topology change. We quickly review canonical quantization of a scalar field in ordinary Minkowski spacetime before moving on to address the procedure in curved spacetime. The treatment given to these subjects closely follows Birrell and Davies \cite{9780511622632}. Then, upon motivating the use of the Sorkin-Johnston (SJ) state as a preferred vacuum state, it will be defined first in terms of a theory's Green's functions and then via Bogolyubov transformations from an arbitrary, orthogonal and complete Klein-Gordon solution set. This will position us to address in the next chapter SJ state construction on spacetimes that are less than amenable to its original definition via Green's functions.

\section{Free Scalar Quantum Field Theory in Flat Spacetime}

Early attempts in the history of quantum theory to construct a relativistically covariant, phenomenologically relevant single particle wave equation with a well defined probabilistic interpretation were quickly riddled with inconsistencies. Among the myriad difficulties encountered were artifacts such as the existence of infinite ``seas'' of negative energy states, non-positive definite probability densities, and the inability to cope, in general, with systems with variable number of particles \cite{AlvarezGaume:2005qb}. As such, the portrait of the dynamics of fundamental particles through the lens of the probabilistic wavefunction was quickly abandoned in favor of quantization of the classical field itself rather than merely its coordinates. The view we take then is that of ``Second Quantization.''

Let $\phi(x) \in \mathbb{R}$ be a spinless massive scalar particle with mass parameter, $m$, such as might describe the post-symmetry breaking Standard Model Higgs degree of freedom \cite{0034-4885-52-4-001}, which satisfies the Klein-Gordon field equation:
\begin{equation}(\partial_{\mu}\partial^{\mu}+m^2)\phi(x)=0.\end{equation}
The field equation can be derived as a result of demanding that the action:
\begin{equation} \label{eq:minklag}
S=\int d^nx \mathcal{L} = \int d^nx \frac{1}{2}(\partial_{\mu}\phi(x) \partial^{\mu}\phi(x)-m^2\phi^2(x))\end{equation}
be invariant under infinitesimal deformations of the field:
\begin{equation}\delta S=0.\end{equation}
That the Minkowski metric in $n$ dimensions,
\begin{equation}ds^2=\eta_{\mu \nu}dx^{\mu}dx^{\nu}=dt^2-\sum_{i=1}^{n-1}(dx^i)^2\end{equation}
has a timelike Killing vector field, $\partial_t \eta_{\mu \nu}=0$, indicates a particular class of solutions to the Klein-Gordon equation:
\begin{equation}u_{\textbf{k}}(x)=N_{\textbf{k}}e^{-ik_{\mu}x^{\mu}}=N_{\textbf{k}}e^{-i(\omega_{\textbf{k}}t-\textbf{k}\cdot\textbf{x})},\end{equation}
where $\omega_{\textbf{k}}>0$ must satisfy $\omega_{\textbf{k}}=\sqrt{\textbf{k}^2+m^2}$. These are known as the positive frequency modes, and are taken together with the negative frequency solutions:
\begin{equation}\bar{u}_{\textbf{k}}(x)=\bar{N}_{\textbf{k}}e^{ik_{\mu}x^{\mu}}=\bar{N}_{\textbf{k}}e^{i(\omega_{\textbf{k}}t-\textbf{k}\cdot\textbf{x})}.\end{equation}

Of course, Minkowski space possesses the full Poincar\'e isometry group. So, there is nothing particularly special about $\partial_t$ other than the fact that it allows us to conveniently and intuitively foliate spacetime into hypersurfaces, $\Sigma_t$, normal to it and define a bilinear form thereon, which is also invariant in time:
\begin{equation}\left(f,g\right)_{KG}=i\int_{\Sigma_t}d^{n-1}x\bar{f} \, \overset\leftrightarrow{\partial_t} \, g=i\int_{\Sigma_t}d^{n-1}x \{\bar{f} (\partial_t g) -  (\partial_t \bar{f})g \}.\end{equation}
This also results in the relativistic normalization constraint:
\begin{equation}N_{\textbf{k}}=\frac{1}{\sqrt{2\omega_{\textbf{k}}(2\pi)^{n-1}}},\end{equation}
which enforces the following on the solution set $\{u_{\textbf{k}}, \bar{u}_{\textbf{k}} \}$:
\begin{equation}\left(u_{\textbf{k}},u_{\textbf{k'}}\right)_{KG}=\delta^{(n-1)}(\textbf{k}-\textbf{k'})\end{equation}
\begin{equation}\left(\bar{u}_{\textbf{k}},\bar{u}_{\textbf{k'}}\right)_{KG}=-\delta^{(n-1)}(\textbf{k}-\textbf{k'})\end{equation}
\begin{equation}\left(u_{\textbf{k}},\bar{u}_{\textbf{k'}}\right)_{KG}=0.\end{equation}
We can then expand any field configuration in terms of linear combinations of these modes, in the classical case, with complex coefficients $a_{\textbf{k}}$ and $\bar{a}_{\textbf{k}}$:
\begin{equation}\phi(x)=\int d^{n-1}k (a_{\textbf{k}}u_{\textbf{k}}(x)+\bar{a}_{\textbf{k}}\bar{u}_{\textbf{k}}(x)).\end{equation}

In order to make the transition to the quantum theory, we elevate the status of our field to an operator valued distribution (this will be defined more precisely in section 2.4), which operates on the Fock space of multiparticle states\footnote{More accurately, the irreducible, symmetric Fock representation of the Hilbert space, $\mathcal{H}$.}, and impose equal time commutation relations. With the scalar Lagrangian density \eqref{eq:minklag}, the conjugate momentum to $\phi$ is:
\begin{equation}\pi(t,\textbf{x})=\frac{\partial \mathcal{L}}{\partial (\partial_t \phi)}=\dot{\phi}(t,\textbf{x}).\end{equation}
The equal time commutation relations are then:
\begin{equation}\left[\phi(t,\textbf{x}),\pi(t,\textbf{y})\right]=i \delta^{(n-1)}(\textbf{x}-\textbf{y})\end{equation}
\begin{equation}\left[\phi(t,\textbf{x}),\phi(t,\textbf{y})\right]=0\end{equation}
\begin{equation}\left[\pi(t,\textbf{x}),\pi(t,\textbf{y})\right]=0.\end{equation}
These are completely equivalent to the following commutation relations imposed upon the complex coefficients, $a_{\textbf{k}}$ and $\bar{a}_{\textbf{k}}$, now elevated to Fock space annihilation and creation operators, $a_{\textbf{k}}$ and $a^{\dagger}_{\textbf{k}}$,
\begin{equation}\left[a_{\textbf{k}},a^{\dagger}_{\textbf{k'}}\right]=\delta^{(n-1)}(\textbf{k}- \textbf{k}')\end{equation}
\begin{equation}\left[a_{\textbf{k}},a_{\textbf{k'}}\right]=0\end{equation}
\begin{equation}\left[a^{\dagger}_{\textbf{k}},a^{\dagger}_{\textbf{k'}}\right]=0,\end{equation}
with the vacuum state being defined as that, which is annihilated by every $a_{\textbf{k}}$:
\begin{equation}a_{\textbf{k}}\left|0\right>=0.\end{equation}
Because it bears heavily upon what follows, it will be useful to examine the Poincar\'e invariance of the vacuum. Let: $\tilde{x}^{\mu} = \Lambda^{\mu}_{\nu} \, x^{\nu} + a^{\mu}$ be a Poincar\'e transformation on the coordinate system corresponding to the vacuum state above, with $\Lambda^{\mu}_{\nu} \in SO(1,3)$ and $a^{\mu}$ a spacetime translation. Note first that:
\begin{equation}
a_{\textbf{k}}\left|0\right>=0 \implies \int d^{n-1}k \frac{1}{\sqrt{2\omega_{\textbf{k}}(2\pi)^{n-1}}} e^{-ik_{\mu}x^{\mu}} a_{\textbf{k}}\left|0\right>=0.
\end{equation}
Now, set ${a^{\mu}=0}$ and consider just Lorentz transformations. Then, under $x \rightarrow \tilde{x}$, 
\begin{equation}
\left|0\right> \rightarrow \left|\tilde{0}\right>,
\end{equation}
and
\begin{equation}
k_{\mu}x^{\mu} \rightarrow \tilde{k}_{\mu}\tilde{x}^{\mu}=k_{\alpha}\Lambda^{\alpha}_{\mu}\Lambda^{\mu}_{\nu}x^{\nu} = k_{\alpha}\delta^{\alpha}_{\nu}x^{\nu}=k_{\mu}x^{\mu},
\end{equation}
while,
\begin{equation}
\int \frac{d^{n-1}k}{2\omega_{\textbf{k}}(2\pi)^{n-1}}
\end{equation}
and
\begin{equation}
\sqrt{{2\omega_{\textbf{k}}(2\pi)^{n-1}}}a_{\textbf{k}}
\end{equation}
are the Lorentz invariant integration measure and annihilation operator, respectively \cite{Srednicki:1019751}. Therefore, $a_{\textbf{k}}\left|\tilde{0}\right>=0,$ which by definition means that under Lorentz transformations, $\left|0\right> = \left|\tilde{0}\right>.$ Meanwhile, allowing $a^{\mu}$ to be nonzero while fixing $\Lambda^{\mu}_{\nu} = \delta^{\mu}_{\nu},$ and noting that wave vectors transform trivially under translations implies that these also are a symmetry of the vacuum.

%%Perhaps include something about now Wick theorem, positivity of Tmn, particle number, and correlators are all now well defined

\section{Scalar Quantum Field Theory in Curved Spacetime}
We now generalize the flat, Minkowski metric, $\eta_{\mu \nu}$ to $g_{\mu \nu}$, the components of the symmetric, nondegenarate, bilinear form $g(\cdot,\cdot)$ on an n-dimensional, smooth Lorentzian manifold M, with signature $(+-\cdot \cdot \cdot \, -)$. Writing $g_{\mu \nu}$ as an $n \times n$ symmetric matrix, let: $g(x)=det(g_{\mu \nu}(x))$ in a coordinate chart, $\{x^{\mu}\}$. Then the Lagrangian density for the real scalar field coupled to gravity is:
\begin{equation}\mathcal{L}=\frac{1}{2}\sqrt{-g(x)}\{g^{\mu \nu}(x) \partial_{\mu} \phi(x) \partial_{\nu} \phi(x) - \left(m^2+ \xi R(x)\right)\phi^2(x)\},\end{equation}
where $R(x)$ is the Ricci scalar and $\xi$ is the dimensionless, numerical coupling of the scalar field to gravity. $\xi = 0$ is called minimal coupling while $\xi=\frac{1}{4}[(n-2)/(n-1)]$ corresponds to conformal coupling. Minimizing the corresponding action with respect to variations in $\phi$,
\begin{equation}\delta S=\delta \int d^nx \mathcal{L} = 0,\end{equation}
yields the field equation:
\begin{equation} \label{eq:curveom}
\left(\Box + m^2 + \xi R(x)\right)\phi(x)=0,
\end{equation}
where,
\begin{equation}\Box \phi(x) = g^{\mu \nu} \nabla_{\mu} \nabla_{\nu} \phi(x) = \frac{1}{\sqrt{-g}} \partial_{\mu} \left[\sqrt{-g} g^{\mu \nu} \partial_{\nu} \phi(x) \right].\end{equation}
In order to perform a well defined quantization of the field theory on $\left(M, g_{\mu \nu}\right),$ we specify that the spacetime be globally hyperbolic. This ensures that $M$ can be foliated into spacelike Cauchy hypersurfaces, $\Sigma$, along a timelike coordinate, and that once a set of conditions for $\phi$
is initialized upon one of the $\Sigma$, its value is determined uniquely on all of $\left(M, g_{\mu \nu}\right)$ through \eqref{eq:curveom}.

The generalization of the Klein-Gordon inner product is
\begin{equation}\left(f,g\right)_{KG}=i \int_{\Sigma} d \Sigma \, n^{\mu} \bar{f} \, \, \overset\leftrightarrow{\nabla}_{\mu} \, g\end{equation}
where
\begin{equation}d \Sigma = d^{n-1}x \sqrt{-g_{\Sigma}(x)}\end{equation}
is the volume form integration measure on the hypersurface $\Sigma$, $n^{\mu}$ are the components of the future-directed, timelike normal vector, orthogonal to $\Sigma$, and $g_{\Sigma}(x)$ is the determinant of the induced metric on $\Sigma$. The Klein-Gordon inner product is invariant across timelike translations. That is, it is independent of the $\Sigma$ upon which it is calculated. As such, by existence and uniqueness theorems for hyperbolic differential equations on $C^{\infty},$ globally hyperbolic manifolds, a complete solution set to the field equation exists, $\{u_i,\bar{u}_i\}$, and once its elements are found to satisfy:
\begin{equation} \label{eq:KGO1}
\left(u_i,u_j\right)_{KG}=\delta_{i j}
\end{equation}
\begin{equation} \label{eq:KGO2}
\left(\bar{u}_i,\bar{u}_j\right)_{KG}=-\delta_{i j}
\end{equation}
\begin{equation} \label{eq:KGO3}
\left(u_i,\bar{u}_j\right)_{KG}=0,
\end{equation}
on any one of the Cauchy hypersurfaces, they will remain so orthonormalized forever. Here, the Roman subscripts are generalized labels to delineate between different modes. For example, $i$ plays the role of the wavenumber $k$ in the case of Minkowski spacetime, or takes on discrete values if spatial sections of the spacetime are compact. Once such a set of modes is found, we can expand the quantized field by analogy with the Minkowski case:
\begin{equation}\phi(x)=\sum_{i} (a_iu_i(x)+{a}^{\dagger}_i\bar{u}_i(x)),\end{equation}
where,
\begin{equation} \label{eq:curvcom1}
\left[a_i,a^{\dagger}_j\right]=\delta_{i j}
\end{equation}
\begin{equation} \label{eq:curvcom2}
\left[a_i,a_j\right]=0
\end{equation}
\begin{equation} \label{eq:curvcom3}
\left[a^{\dagger}_i,a^{\dagger}_j\right]=0,
\end{equation}
and define a vacuum by:
\begin{equation}a_i\left|0\right>=0 \; \forall \, i.\end{equation}
The operative word above is ``a.'' In the absence of a high degree of symmetry of the spacetime\footnote{The most desirable situation is when the spacetime admits a global timelike Killing vector field, $\kappa=\frac{d}{dt}$, which commutes with the Klein-Gordon operator, $\left(\Box + m^2 + \xi R(x)\right)$. In this case, a solution set can be found, which is positive frequency with respect to the Killing vector, $\kappa \, u_k(x) = -i \omega_k u_k(x)$ and the corresponding vacuum will be unique.}, the solution set $\{u_i,\bar{u}_i\}$ will not be unique, and the number of corresponding vacua will be as multitudinous as the different sets available with which to decompose $\phi$ \cite{Wald:2009uh}. The fundamental idea in general relativity, that physics occurs on $M$ rather than in the coordinate charts, finds its expression in semiclassical gravity by rendering the vacuum generally meaningless. In particular, if we let $\{v_i,\bar{v}_i\}$ be another Klein-Gordon orthonormal solution set, then the field can be expanded as:
\begin{equation}\phi(x)=\sum_{i} (b_iv_i(x)+{b}^{\dagger}_i\bar{v}_i(x)),\end{equation}
where now the vacuum is defined by:
\begin{equation}b_i\left|\tilde{0}\right>=0 \; \forall \, i,\end{equation}
and the two sets are related by a Bogolyubov transformation:
\begin{equation} \label{eq:BOGOLYUBOV}
v_i(x)=\sum_j \left(A_{i j}u_j(x) + B_{i j}\bar{u}_j(x)\right).
\end{equation}
With the Klein-Gordon orthonormalization conditions, the Bogolyubov coefficients can be obtained by:
\begin{equation}A_{i j}=\left(v_i,u_j\right)_{KG} \; \; \; \; \; B_{i j}=-\left(v_i,\bar{u}_j\right)_{KG}\end{equation}
Equating then, the two field expansions,
\begin{equation}\phi(x)=\sum_{i} (a_iu_i(x)+{a}^{\dagger}_i\bar{u}_i(x))=\sum_{i} (b_iv_i(x)+{b}^{\dagger}_i\bar{v}_i(x))\end{equation}
$$=\sum_{i j} \left( (b_iA_{i j}+b^{\dagger}_i \bar{B}_{i j})u_j(x) + (b_iB_{i j}+b^{\dagger}_i \bar{A}_{i j})\bar{u}_j(x)\right),$$
and taking  the Klein-Gordon inner products again, once with $u_{k}(x)$ and once with $\bar{u}_{k}(x),$ yields the transformation properties of the operator coefficients:
\begin{equation} a_{k}=\sum_j \left( b_jA_{j k}+b^{\dagger}_j \bar{B}_{j k}\right) \end{equation}
\begin{equation} a^{\dagger}_{k}=\sum_j \left(b_jB_{j k}+b^{\dagger}_j \bar{A}_{j k}\right).\end{equation}
That the vacuum is a generally variant quantity can then be seen by calculating:
\begin{equation}\left<\tilde{0}\right| a^{\dagger}_{k}a_{k} \left|\tilde{0}\right> = \left<\tilde{0}\right| \sum_i \left(b_iB_{i k}+b^{\dagger}_i \bar{A}_{i k}\right) \sum_j \left( b_jA_{j k}+b^{\dagger}_j \bar{B}_{j k}\right) \left|\tilde{0}\right> =\sum_i \left|B_{i k}\right|^2,\end{equation}
which of course does not vanish unless each $B_{i k}$ does. This non-uniqueness of the vacuum has led to concrete physical predictions such the Unruh effect \cite{Unruh:1976db}, Hawking radiation \cite{Hawking:1974sw}, and the production of Gaussian-distributed random perturbations within the theoretical framework of cosmic inflation \cite{1992PhR...215..203M}. However, the lack of even a preferred ground state, to say nothing of a unique one, presents a serious ontological quandary to anyone attempting to retain the familiar notion of quantal excitations as particles. It would seem pertinent then to try and identify either some criterion by which to recognize such a preferred vacuum or methods by which to construct one. In the next section, we will present the fundamental ideas of causal set theory with a view towards using the closely associated Sorkin-Johnston state as a natural option.

For completeness, we include the remaining transformation properties of the field modes, operator coefficients, and Bogolyubov coefficients as will be relevant for our discussion later on. The inverse transformation from the $\{v_i,\bar{v}_i\}$ to the $\{u_i,\bar{u}_i\}$ is:
\begin{equation}
u_i(x)=\sum_j \left( \bar{A}_{j i}v_j(x) - B_{j i}\bar{v}_j(x) \right),
\end{equation}
while for the operator coefficients, it is
\begin{equation}
b_i=\sum_j \left( \bar{A}_{i j}a_j - \bar{B}_{i j}a^{\dagger}_j \right).
\end{equation}
Finally, preservation of \eqref{eq:KGO1}-\eqref{eq:KGO3} under the transformation requires that:
\begin{equation} \label{eq:Bogid1}
\sum_k \left( A_{i k} \bar{A}_{j k} - B_{i k} \bar{B}_{j k} \right) = \delta_{i j},
\end{equation}
\begin{equation} \label{eq:Bogid2}
\sum_k \left( A_{i k} B_{j k} - B_{i k} A_{j k} \right) = 0.
\end{equation}

\section{Causal Set Theory}
%%Not exactly sure what to do here, but probably talk about causal set basics, construct discrete retarded green's function on causal set, P-J function, SJ state...

Causal set theory is a framework that proposes to rectify the three ``infinities'' of modern theoretical physics, namely, the ultraviolet divergences arising in relativistic quantum field theory\footnote{The most egregious example here, of course, is the perturbative non-renormalizability of interactions involving the graviton. \cite{'tHooft:1978id}} \cite{Srednicki:1019751}, the curvature singularities of General Relativity \cite{Wald:1984rg}, and the unbounded result when the Bekenstein-Hawking entropy is calculated directly by way of quantum gravitational degrees of freedom \cite{PhysRevD.34.373}. The basic idea that spacetime should be in some way discrete is common to many of the approaches aimed at successfully quantizing gravity \cite{'tHooft:1978id}. The causal sets program shares this view and adopts discreteness along with causality as its fundamental kinematic axioms \cite{PhysRevLett.59.521}.

In fact, given only the causal relations between the elements of a set of discrete points in a given manifold, all of the topological information and nearly all of the metrical information of the host manifold can be recovered, in addition to its dimension and differential structure \cite{:/content/aip/journal/jmp/17/2/10.1063/1.522874} \cite{:/content/aip/journal/jmp/18/7/10.1063/1.523436}. Then, the only piece of information missing is an overall conformal factor of the metric, which can be naturally fixed by requiring that the number of causal set elements (at some embedding density) corresponds to the volume of the continuum spacetime region under consideration. Ongoing formulations have then typically involved a sum over histories or quantum measure theoretic approach to the dynamics on the space of all causal sets \cite{Henson:2006kf} \cite{doi:10.1142/S021773239400294X}. The eventual goal is to discard the concept of a manifold entirely in favor of the causal set as the most fundamental structure. Interestingly, even without a full quantum dynamics in place, the effects of this type of discrete geometry have already been shown to be phenomenologically prescient as regards, most notably, the minscule value of the cosmological constant \cite{Sorkin:2007bd}. For our purposes however, we will employ the causal set as a Lorentz invariant discretization of spacetime, upon which it is convenient, for example, to perform semiclassical calculations numerically. To make this all concrete, we introduce a definition.

A \textit{causal set}, denoted by $\left(\mathcal{C}, \prec \right),$ is a set $\mathcal{C}$ endowed with a locally finite partial order relation $\prec$, which satisfies the following properties:
\vspace{5mm}

1.) Transitivity: \, \, \, \, \,$x \prec y$ and $y \prec z \implies x\prec z$ \, \, \, \, \, \, $\forall \, \, x,y,z \in \mathcal{C}$

2.) Irreflexivity: \, \, \, \, \, $x \nprec x$ \, \, \, \, \, \, \, \, \, \, \, \, \, \, \, \, \, \, \, \, \, \, \, \, \, $\forall \, \, x\in \mathcal{C}$

3.) Local Finiteness: \, $\left| \{ y \in \mathcal{C} \, | \, x \prec y \prec z \} \right| < \infty$ \, \, \, \, \, \, \, \,$\forall \, \, x,y,z \in \mathcal{C}.$
\vspace{5mm}

To borrow the nomenclature of global causal analysis, we say that ``$x$ is in the causal past of $y$'' if $x \prec y$. In addition, we write $x \preceq y$ in the case that $x \prec y$ or $x=y$. It is illustrative to consider these conditions within the context of embedding in a continuum manifold (while keeping in mind that we should then refocus on the causal set). Transitivity formalizes the intuitive idea that if we can establish timelike or null worldlines from $x$ to $y$ and from $y$ to $z$, then we ought also to be able to establish one from $x$ to $z$. Meanwhile, irreflexivity prohibits the existence of the discrete equivalent of closed timelike curves of the form $x \prec y \prec z \prec \cdot \cdot \cdot \prec w \prec x.$ In other words, $x$ cannot be in its own causal past. Finally, the condition that the causal set be locally finite is how one specifies that its elements are discrete. That is, if we look at the region marked out by the intersection of the past lightcone of $z$ and the future lightcone of $x$, the cardinality of the causal subset comprised of elements therein should be finite.

Now, given an n-dimensional Lorentzian manifold $(M,g_{\mu \nu}),$ we can construct a causal set $\left(\mathcal{C}, \prec \right)$ whose partial order relation bears the geometrical information of the parent manifold by way of a process called \textit{sprinkling}. While discretization of a manifold might seem to be most easily carried out by simply laying down a rectangular, or other regular, lattice whose elements obey the causal precedence relations of the manifold, specialization to Minkowski space and consideration of a boost along some axis quickly shows that this will not be Lorentz invariant\footnote{Inspect for example the effect of length contraction and time dilation on a spacetime square of side length $\ell$.}. However, distribution of points into Minkowski space by way of a Poisson process does respect Lorentz symmetry.

Putting aside for a moment the dire problem of quantum gravity, consider some stochastic process such as children dropping their ice cream cones at the beach, which occurs along a homogenous time function $\tau$ with an average rate of $\omega.$ The probability that there will be $n$ instances of this happening in some time interval $\Delta \tau$ can then be modeled by the Poisson distribution \cite{Poisson1837}:

\begin{equation}
P(n)=\frac{e^{-\omega \Delta \tau}(\omega \Delta \tau)^{n}}{n!}.
\end{equation}
The generalization to the probability of $n$ elements being counted within a spacetime volume $V$ given an average point embedding density of $\rho$ is\footnote{In accordance with the idea that distance measures will likely be meaningless below roughly the Planck length ($\ell_p=1.6 \times 10^{-33}$ cm) in any final quantum theory of gravity, it is typically taken to be the case that $\rho \sim \ell_p^{-n}$ \cite{Henson:2006kf}.}:

\begin{equation} \label{eq:Poisson}
P(n)=\frac{e^{-\rho V}(\rho V)^{n}}{n!}.
\end{equation}
Since the expectation value of the distribution, $\left<N\right> =\rho V$ is clearly Lorentz invariant, so too is $P(n)$, as will be the embedding predicated thereon, but in a statistical sense. This is what is meant by a sprinkling- a discretization of the manifold whose elements obey the counting statistics of \eqref{eq:Poisson}. If in the limit $\rho \rightarrow \infty$, the original manifold $M$ is recovered, the embedded causal set $\mathcal{C}$ will be called a \textit{faithful embedding}. In practice, the statistical Lorentz invariance means that in order to derive meaningful physical quantities, one can use a garden variety random number generator computationally to discretize $M$, but will need to average results over a number of such discretizations.

%%Poisson variations as conjugate to cosmological constant?! Footnote? Appendix (w/ Einstein-Hilbert action?)

Once such a causal set has been thus obtained, a natural next step is to see how scalar quantum field theory might be affected by this discreteness. While still not fully quantum gravitational, this sort of analysis could lead to the extension of the semiclassical causal set phenomenology\footnote{While tangential for this work, the heuristic prediction of $\Lambda$ is interesting. It goes as follows. Assume $\left<\Lambda \right>=0$ and what actually drives acceleration is its quantum fluctuation, $\Delta \Lambda.$ Identify $\Lambda$ and $V$ as conjugate quantities as viewed by unimodular gravity in the Einstein-Hilbert action, $-\Lambda \int d^4x \sqrt{-g}=-\Lambda V$. They then satisfy an uncertainty relation: $\Delta \Lambda \Delta V \sim 1$. In natural units, $\hbar=c=G=1 \implies \ell_p=1 \implies \rho=1$ in \eqref{eq:Poisson}. But $N \sim \left<N\right> \pm \Delta N \sim V \pm \sqrt{V},$ using the mean and standard deviation of the Poisson distribution. Then, $V \sim N \pm \sqrt{V} \sim \left<V\right> \pm \Delta V \implies \Delta V= \sqrt{V}$, so that $\Delta \Lambda \sim V^{-\frac{1}{2}}$. Taking $V \sim H^{-4}$, where $H \sim 10^{-61}$ is the Hubble constant in natural units, $\Delta \Lambda \sim 10^{-122}$.} of the type found in \cite{Ahmed:2002mj} \cite{Ahmed:2012ci}. Discrete, Lorentz invariant d'Alembertian operators have been found for flat spacetimes of arbitrary dimension $n$ \cite{Dowker:2013vba}. So, we could in principle use these to find sets of discrete eigenmodes of the Klein-Gordon field. We will follow a different route to determine these modes based on the field's discrete retarded propagator on the causal set. To do this, we will need the $p \times p$ causal adjacency matrix $C$ whose entries are defined by:

\begin{equation} \label{eq:causalmatrix}
C_{i j}:= \left \{
\begin{tabular}{cc}
$1$ & if $x_i \prec x_j$ \\
$0$ & otherwise,
\end{tabular}
\right .
\end{equation}
where $x_i, x_j \in \mathcal{C}$ and $p$ is the cardinality of $\mathcal{C}$. The causal matrix encodes all of the binary order relations within the set. Later, we'll see how this relates to the field's discrete retarded propagator, but first, we review some aspects of Green's functions in the continuum theory.
%%while d'alambertian has been found, techniques of solution of modes with is not trivial. Let's look at continuum green's functions and see if we can find some way to derive modes without this. -> possible to find green's function without wave equation! (at least in 2-d). causal matrix here or there?

%%Go through causal diamond example continuum -> discrete solving deq to show, in addition to original derivation of green's function, the idea that Gr is just constant with causal past. Then, later can talk about numerical determination on the cone with images (Schmitzer's thesis? Princeton astrophysics paper?)

\section{Green's Functions and the Sorkin-Johnston State}
%%Talk about Wightman Axioms Here?
In our discussion of canonical quantization of the scalar field in curved spacetime (Section 2.2), recall that we stipulated commutation relations for the creation and annihilation operators in the expansion of the field, \eqref{eq:curvcom1}-\eqref{eq:curvcom3}. However, nothing was said about the corresponding form of the \textit{field} commutators as we had for Minkowski space. It turns out that they can be stated in fully covariant form \cite{1952}. In order to do this, calculate the vacuum expectation value (VEV) of the field commutator in some mode set:
\begin{equation}
\left<0\right| \left[\phi(x),\phi(y) \right] \left|0\right>=\sum_i \left( u_i(x)\bar{u}_i(y) - \bar{u}_i(x)u_i(y) \right).
\end{equation}
Then, defining the Pauli-Jordan Function, $\Delta(x,y)$, as
\begin{equation} \label{eq:PauliJ}
i\Delta(x,y):=\sum_i \left( u_i(x)\bar{u}_i(y) - \bar{u}_i(x)u_i(y) \right),
\end{equation}
we can write
\begin{equation} \label{eq:covcom}
\left[\phi(x),\phi(y) \right]=i\Delta(x,y) \, \mathbb{I},
\end{equation}
where $\mathbb{I}$ is the Fock space identity operator. From now on, we will drop $\mathbb{I}$ under the understanding that its existence is implicit when dealing with Fock states. Since the left hand side of \eqref{eq:covcom} bears no indication of the coordinate basis with which the field is decomposed, $i\Delta(x,y)$ had better not either. Luckily, so long as \eqref{eq:Bogid1} and \eqref{eq:Bogid2} are satisfied, such as between two bases, which are correctly orthonormalized under $(\cdot,\cdot)_{KG},$ the Pauli-Jordan function indeed is also basis independent \cite{Afshordi:2012jf}.

Moreover, we now make precise the sense in which $\phi$ is an operator valued \textit{distribution} in the general case. Given some smooth functions of compact support on M, $f,h \in C^{\infty}_0(M),$ the rigorous expression of \eqref{eq:covcom} is:
\begin{equation} \label{eq:covcomdist}
\left[\phi(f),\phi(h) \right]=i \int_M d^nx \, d^ny \sqrt{-g(x)} \sqrt{-g(y)} f(x) \Delta(x,y) h(y).
\end{equation}
Additionally, expressions such as: $\phi(x) \left|\psi \right>$ should formally read
\begin{equation} \label{eq:opdist}
\phi(f) \left|\psi \right> = \int_M d^nx \sqrt{-g(x)} \phi(x) f(x) \left|\psi \right>,
\end{equation}
for all $f$. The requirement that field operators be defined as distributional addresses ultraviolet divergence issues arising from the ill-definition of a field at a point by ``smearing'' with $f$ and ensures that after application to Fock space kets, the states are still normalizable \cite{2011arXiv1110.5013C}. In practice, equations such as \eqref{eq:opdist} are most relevant in axiomatic quantum field theory (see for example \cite{streater}), which will not be the approach most closely followed here. More importantly for us, equation \eqref{eq:covcomdist} suggests the use of the Pauli-Jordan function as a bilinear integral operator. As such, define:
\begin{equation}
\left(i\Delta f \right)(x):=i\int_M d^ny \sqrt{-g(y)} \Delta(x,y) f(y).
\end{equation}
As can be seen from its definition in \eqref{eq:PauliJ}, $i\Delta(x,y)$ enjoys two important properties as an operator, namely:
\vspace{5mm}

1.) Skew-symmetry: \, \, $i\Delta(y,x)=-i\Delta(x,y)$

2.) Hermiticity: \, \, $(i\Delta(x,y))^{\dagger}=i\Delta(x,y).$
\vspace{5mm}

The second property means that if:

\begin{equation} \label{eq:Lsquaredin}
\left<f,h\right>=\int_M d^nx \sqrt{-g(x)} \, \bar{f}(x)h(x)
\end{equation}
is the $L^2(M)$ inner product, then $i\Delta(x,y)$ is self-adjoint on $L^2(M)$. That is:
\begin{equation}
\left<f,i\Delta h\right>=\left<i\Delta f,h\right>,
\end{equation}
and by the spectral theorem, $i\Delta(x,y)$ can be diagonalized. This object then becomes the linchpin for the Sorkin-Johnston proposal for a preferred vacuum state \cite{Johnston:2009fr} \cite{Sorkin:2011pn}.

In the canonical formulation of quantum field theory, Wick's Theorem allows us to write any time-ordered product of fields as a polynomial in Wightman two-point functions \cite{Srednicki:1019751}. In a given Klein-Gordon solution set $\{u_i,\bar{u}_i\}$, the Wightman function can be written as:
\begin{equation}
W(x,y) = \left<0\right| \phi(x) \phi(y) \left|0\right> = \sum_i u_i(x) \bar{u}_i(y).
\end{equation}
Notice that, in contrast to the Pauli-Jordan function, the Wightman function is not in general invariant under Klein-Gordon norm preserving Bogolyubov transformations. So long as the vacuum state can be taken to be Gaussian, specifying which Wightman function to use in the Wick polynomials is exactly equivalent to choosing a mode set in which to expand $\phi$. Now, since $i\Delta(x,y)$ is diagonalizable and Hermitian, it has a set of $L^2(M)$ orthonormal eigenfunctions, $\{u_a\}$ with real eigenvalues, $\{\lambda_a\}$ such that:
\begin{equation} \label{eq:PJeigen}
(i\Delta u_a)(x) = \lambda_a u_a(x).
\end{equation}
Taking the complex conjugate of \eqref{eq:PJeigen}, and noting that $\Delta(x,y)$ is a real valued kernel and $\lambda_a \in \mathbb{R} \, \, \forall \, \, a$,
\begin{equation}
(i\Delta \bar{u}_a)(x) = -\lambda_a \bar{u}_a(x).
\end{equation}
Evidently, the eigenvalues come in positive-negative pairs. We can now decompose $i\Delta(x,y)$ in the $\{u_a, \bar{u}_a\}$ modes:
\begin{equation}
i\Delta(x,y) = \sum_a \lambda_a \left(u_a(x)\bar{u}_a(y) - \bar{u}_a(x) u_a(y) \right).
\end{equation}
The first and second terms are referred to respectively as the positive and negative parts of the decomposition. If we now define
\begin{equation}
u^{SJ}_a(x) := \sqrt{\lambda_a} u_a(x), 
\end{equation}
then the Sorkin-Johnston vacuum is obtained by expanding our field in terms of these modes,
\begin{equation}
\phi(x) = \sum_b \left(a_b u^{SJ}_b(x) + a^{\dagger}_b \bar{u}^{SJ}_b(x) \right),
\end{equation}
and using the corresponding coefficients as the correct Fock space creation and annihilation operators:
\begin{equation}
a_b \left|SJ \right> = 0 \, \, \forall \, \, b.
\end{equation}
The Wightman two-point function in this vacuum is:
\begin{equation}
\begin{aligned}
W^{SJ}(x,y) &= \left<SJ\right| \phi(x) \phi(y) \left|SJ \right> \\ &= \sum_a u^{SJ}_a(x) \bar{u}^{SJ}_a(y) \\ &= \sum_a \lambda_a u_a(x)\bar{u}_a(y) \\ &= Pos\left[ i\Delta(x,y) \right].
\end{aligned}
\end{equation}
In adopting this two-point function, we specify our full quantum field theory through Wick's Theorem. If one considers that $i\Delta(x,y)$ encodes everything that is ``quantum'' in a theory via the commutator, and does so in a fully covariant way, thereby satisfying also the relativistic notion that coordinates are unimportant, it seems reasonable to think that imprinting this information upon the vacuum state via diagonalization in the expansion basis should select the most physically relevant of the vacua available.

We would like to check that after convolution with the SJ Wightman function, the projection of a test function $f$ onto itself is positive-semidefinite. In other words, $W^{SJ}$ should be positive as should any reliable two point function. This follows from the fact that using the SJ modes is just another way of decomposing the field:

\begin{equation}
\begin{aligned}
\left<f,W^{SJ}f\right>&=\int_M \int_M d^nx d^ny \sqrt{-g(x)} \sqrt{-g(y)} \bar{f}(x) W^{SJ}(x,y) f(y)
\\&=\int_M \int_M d^nx d^ny \sqrt{-g(x)} \sqrt{-g(y)} \bar{f}(x) \left<SJ\right|\phi(x)\phi(y)\left|SJ\right> f(y)
\\&=\left<SJ\right| \int_M \int_M d^nx d^ny \sqrt{-g(x)} \sqrt{-g(y)} \bar{f}(x) \bar{\phi}(x)\phi(y) f(y) \left|SJ\right>
\\&= \left<\phi(f)|\phi(f)\right>,
\end{aligned}
\end{equation}
where the reality of $\phi$ has been used. So long as the fields satisfy the Wightman axioms, this Hilbert inner product is positive-semidefinite. Hence, so is $W^{SJ}$.

\begin{equation} \label{eq:POSITIVITY}
\left<f,W^{SJ}f\right> \geq 0.
\end{equation}
It is also easy to see that we can recover the commutator via:
\begin{equation} \label{eq:commconsist}
\begin{aligned}
W^{SJ}(x,y)-W^{SJ}(y,x)&=W^{SJ}(x,y)-\bar{W}^{SJ}(x,y)\\&=i\Delta(x,y)\\&=\left[\phi(x),\phi(y)\right],
\end{aligned}
\end{equation}
and that the two point function satisfies the Klein-Gordon equation in both arguments:
\begin{equation}
\left(\Box + m^2 + \xi R\right)W^{SJ}=0.
\end{equation}
That $W^{SJ}$ specifies a unique Wightman function is given by \eqref{eq:POSITIVITY} and \eqref{eq:commconsist} along with a criterion which the author of \cite{michelthesis} has termed ``orthogonal supports'':
\begin{equation} \label{eq:orthogsupp}
W^{SJ}\bar{W}^{SJ}:=\int_M d^nz \sqrt{-g(z)} W^{SJ}(x,z)\bar{W}^{SJ}(z,y)=0.
\end{equation}
This can be shown for finite matrix operators quickly. Let $W_1$ and $W_2$ be two such Wightman functions. Then, equation \eqref{eq:commconsist} implies $W_1-\bar{W}_1= W_2-\bar{W}_2$. Squaring both sides gives $(W_1-\bar{W}_1)^2= (W_2-\bar{W}_2)^2$, which implies $(W_1+\bar{W}_1)^2= (W_2+\bar{W}_2)^2$ by \eqref{eq:orthogsupp}. But by \eqref{eq:POSITIVITY} we can take the unique square root of both sides to get $W_1+\bar{W}_1= W_2+\bar{W}_2$. Together with \eqref{eq:commconsist} again this proves that $W_1=W_2$.
%%properties of wightman function (positivity etc.)

Along with the Pauli-Jordan and Wightman functions, we can also define the retarded and advanced Green's functions of the theory:

\begin{equation} \label{eq:RETOTO}
G_R(x,y):= \chi (y \prec x) \Delta(x,y)
\end{equation}
\begin{equation}
G_A(x,y):= -\chi (x \prec y) \Delta(x,y)\footnote{It is evident from these definitions that $G_A(x,y)=G_R(y,x)$ by the skew-symmetry of $i\Delta(x,y)$.},
\end{equation}
$\chi(s)$ being 1 if its argument is logically true and 0 otherwise. $\Delta(x,y)$ can then be recovered by:
\begin{equation} \label{eq:deltadiff}
\Delta(x,y)=G_R(x,y)-G_A(x,y)
\end{equation}
by using the property that $\chi(y \prec x)+\chi(x \prec y)=1$ on the domain over which the commutator \eqref{eq:covcom} is non-vanishing. Applying \eqref{eq:curveom} to \eqref{eq:covcom} implies that:

\begin{equation}
\left(\Box + m^2 + \xi R\right)\Delta(x,y)=0
\end{equation}
in both arguments, and the retarded and advanced propagators obey the Green's function equations:
\begin{equation} \label{eq:AReom}
\left(\Box + m^2 + \xi R\right) G_{R,A}(x,y)= - \frac{\delta^{(n)}(x-y)}{\sqrt{-g}}
\end{equation}
in both arguments as well. Note that we could have begun with \eqref{eq:AReom} and run the argument in reverse in order to construct $i\Delta(x,y)$ through \eqref{eq:deltadiff}. This actually furnishes another way to find the Pauli-Jordan kernel, and has been put to good effect to find the SJ state on the causal diamond in \cite{Afshordi:2012ez}.

A final Green's function, which typically finds employment in curved spacetime is Hadamard's elementary function:

\begin{equation}
\begin{aligned}
G^{(1)}(x,y) &:=\left<0 \right| \{ \phi(x), \phi(y) \} \left|0 \right> \\ &= W(x,y) + \bar{W}(x,y) \\&= \sum_i \left( u_i(x)\bar{u}_i(y) + \bar{u}_i(x) u_i(y)\right).
\end{aligned}
\end{equation}
Since the Pauli-Jordan function is state-independent, the Hadamard function must carry all of the state-dependence that the Wightman function does, as can be seen by the relation:

\begin{equation} \label{eq:wightdeco}
W(x,y)=\frac{i}{2}\Delta(x,y)+\frac{1}{2}G^{(1)}(x,y).
\end{equation}
It should be noted also that although these Green's functions have been discussed within the context of the SJ formalism, \eqref{eq:RETOTO}-\eqref{eq:wightdeco} are valid outside the formalism as well.

%%talk about mathematical subtleties? eigenvalue ranges? properties of W? Hilbert-schmidt/bounded M etc?
%%Definitely talk about other green's functions/construction of idelta from these/hadamard function/that they satisfy KG equation
%%Cite Michel's deSitter paper

\section{The Bogolyubov Method}

In principle, we are now equipped to find the SJ state on any spacetime with which it is compatible\footnote{The spacetime will need to be globally hyperbolic and bounded. Global hyperbolocity has been addressed in section 2.2. The boundedness requirement arises from the fact that $i\Delta$ is an integral operator and is hence highly non-local and potentially divergent. In order to find SJ modes on unbounded spacetimes, integral cutoffs can typically be introduced and then their limits taken to infinity, although there are exceptions where this procedure is ill-defined and the two point function shows sensitivity to the limiting method \cite{Aslanbeigi:2013fga}.}, be it flat, curved, or topologically unconventional. In practice however, the geometry of the spacetime may be such that implementing either the mode sum in \eqref{eq:PauliJ} or finding the retarded and advanced propagators through \eqref{eq:AReom} is prohibitively difficult. In cases such as these it would be advantageous to have an alternate method to find the Sorkin-Johnston modes. Happily, the SJ modes, just like any other orthonormal Klein-Gordon solution set, are also related to each other solution set by a Bogolyubov transformation, and following \cite{Afshordi:2012jf} \cite{Aslanbeigi:2013fga}, we can exploit this fact to find them.

Let $\{u_a,\bar{u}_a\}$ be the desired mode set in which $i\Delta$ is diagonal and $\{v_i,\bar{v}_i\}$ some other arbitrary set, which solves \eqref{eq:curveom} and satisfies \eqref{eq:KGO1}-\eqref{eq:KGO3}. Then,

\begin{equation}
\begin{aligned}
(i\Delta u_a)(x) &=\int d^ny \sqrt{-g(y)} \sum_i \left( v_i(x)\bar{v}_i(y)-\bar{v}_i(x)v_i(y)\right)u_a(y) \\&= \sum_i \left( \left<v_i,u_a \right> v_i(x) - \left<\bar{v}_i,u_a \right> \bar{v}_i(x) \right) \\&= \lambda_a u_a(x)
\end{aligned}
\end{equation}
In order to bring this into the form \eqref{eq:BOGOLYUBOV}, make the definitions $A_{a i} := \frac{\left<v_i,u_a \right>}{\lambda_a}$ and $B_{a i}:=-\frac{\left<\bar{v}_i,u_a \right>}{\lambda_a}$. Then,

\begin{equation} \label{eq:memurr}
u_a(x)= \sum_i \left( A_{a i}v_i(x) + B_{a i} \bar{v}_i(x) \right).
\end{equation}
Taking the $L^2(M)$ inner product of \eqref{eq:memurr}, once with $v_j(x)$ and once with $\bar{v}_j(x)$ yields the two relations for the Bogolyubov coefficients: 

\begin{equation} \label{eq:ayyy}
A_{a j}=\frac{1}{\lambda_a} \sum_i \left( A_{a i} \left<v_j,v_i \right>  + B_{a i} \left<v_j,\bar{v}_i \right>\right)
\end{equation}
\begin{equation} \label{eq:bayyy}
B_{a j}=-\frac{1}{\lambda_a} \sum_i \left( A_{a i} \left<\bar{v}_j,v_i \right>  + B_{a i} \left<\bar{v}_j,\bar{v}_i \right>\right).
\end{equation} 
These, along with \eqref{eq:Bogid1} and \eqref{eq:Bogid2} place a set of four constraints on the coefficients, which when satisfied, will determine the modes $\{u_a,\bar{u}_a\}$ in terms of $\{v_i,\bar{v}_i\}$. In general, this is no less trivial than the previously discussed solution methods for $i\Delta$. Let us assume that whatever solutions we find for the $\{v_i,\bar{v}_i\}$ set are orthogonal under the $L^2(M)$ inner product:

\begin{equation}
\left<v_i,v_j \right> = \left<v_i,v_i\right> \delta_{i j}
\end{equation}
\begin{equation}
\left<v_i,\bar{v}_j \right> = \left<v_i,\bar{v}_{-i}\right> \delta_{i,-j}.
\end{equation}
Like with $i\Delta$, the integrals involved in the $L^2(M)$ inner products may not be finite. In this case, it will similarly be necessary to introduce integral limit regulators and then take them to infinity later on. This postulate greatly simplifies \eqref{eq:ayyy} and \eqref{eq:bayyy}:

\begin{equation}
\begin{aligned}
A_{a j} &=\frac{1}{\lambda_a} \sum_i \left( A_{a i} \left<v_j,v_j \right> \delta_{j i} + B_{a i} \left<v_j,\bar{v}_{-j} \right>\delta_{j, -i} \right) 
\\&= \frac{1}{\lambda_a} \left( A_{a j} \left<v_j,v_j \right> + B_{a -j} \left<v_j,\bar{v}_{-j} \right>\right)
\end{aligned}
\end{equation}
\begin{equation}
\begin{aligned}
B_{a j} &=-\frac{1}{\lambda_a} \sum_i \left( A_{a i} \left<\bar{v}_j,v_{-j} \right> \delta_{j, -i} + B_{a i} \left<\bar{v}_j,\bar{v}_{j} \right>\delta_{j i} \right) 
\\&= -\frac{1}{\lambda_a} \left( A_{a -j} \left<\bar{v}_j,v_{-j} \right> + B_{a j} \left<\bar{v}_j,\bar{v}_{j} \right>\right).
\end{aligned}
\end{equation}
Another simplification comes about, and the consistency of all four constraint equations holds, if we write $A_{a j}=A_a \delta_{a j}$ and $B_{a j}=B_a \delta_{a, -j}$. This says that whatever linear combination of $\{v_i,\bar{v}_i\}$ constitutes the $\{u_a,\bar{u}_a\}$ modes, it will only ever mix modes of the same frequency, up to a sign. The set of equations, which now needs to be solved is:

\begin{equation} \label{eq:TBOG1}
A_a=\frac{1}{\lambda_a} \left(A_a \left<v_a,v_a \right> + B_a \left<v_{a},\bar{v}_{-a} \right> \right)
\end{equation}

\begin{equation} \label{eq:TBOG2}
B_a=-\frac{1}{\lambda_{a}} \left( A_a \left< \bar{v}_{-a},v_a \right> + B_a\left<\bar{v}_{-a} ,\bar{v}_{-a} \right> \right)
\end{equation}

\begin{equation} \label{eq:TBOG3}
\left|A_a \right|^2-\left|B_a \right|^2=\frac{1}{\lambda_a}
\end{equation}

\begin{equation} \label{eq:TBOG4}
A_aB_{-a}-B_aA_{-a}=0.
\end{equation}
The modification to \eqref{eq:Bogid1} in \eqref{eq:TBOG3} is so that the transformation to the SJ modes, rather than the Pauli-Jordan eigenfunctions, will be unitary. By rearranging and dividing equations \eqref{eq:TBOG1} and \eqref{eq:TBOG2} and noting that $\left<\bar{v}_{-a} ,\bar{v}_{-a} \right>=\left<v_{-a} ,v_{-a} \right>$, one finds a quadratic equation whose solution renders the set of eigenvalues:

\begin{equation}
\lambda_a=\frac{1}{2}\left[ \left< v_a,v_a \right> -\left< v_{-a},v_{-a}\right> \pm \sqrt{\left( \left< v_a,v_a \right> +\left<v_{-a},v_{-a} \right>\right)^2 - 4 \left| \left< v_a,\bar{v}_{-a}\right> \right|^2 } \, \right].
\end{equation}
With a view towards the Lorentzian cone under consideration\footnote{See equation \eqref{eq:L21}.} in Chapter 3, we will take $\left<v_a,v_a\right>=\left<v_{-a},v_{-a}\right>$ so that:

\begin{equation} \label{eq:EIGENBALUES}
\lambda_a=\lambda_{-a}=\pm \sqrt{\left|\left<v_a,v_a\right>\right|^2 - \left|\left<v_a,\bar{v}_{-a}\right>\right|^2},
\end{equation}
where now the signs in the eigenvalue solutions just duplicate the parity with which they appear in the solution set. In the expressions below, we take them to be positive for convenience.

Under these conditions, we can now finally state what the Bogolyubov coefficients are \cite{Afshordi:2012jf}:

\begin{equation} \label{eq:Aa1}
A_{a}=\left[\frac{\lambda_a + \left<v_a,v_a\right>}{2\lambda_a^2} \right]^\frac{1}{2}
\end{equation}
\begin{equation} \label{eq:Ba1}
B_{a}=-\left[\frac{-\lambda_a + \left<v_a,v_a\right>}{2\lambda_a^2} \right]^\frac{1}{2} e^{-iArg\left(\left<v_a,\bar{v}_{-a} \right> \right)}.
\end{equation}
They can be re-written slightly as

\begin{equation}
A_{a}=\frac{1}{\sqrt{2\lambda_a}}\left[1+\frac{1}{\sqrt{1-\frac{|\left<v_{a},\bar{v}_{-a}\right>|^2}{\left|\left<v_{a},v_{a}\right>\right|^2}}}\right]^{\frac{1}{2}}
\end{equation}
and
\begin{equation}
B_{a}=-\frac{1}{\sqrt{2\lambda_a}}\left[-1+\frac{1}{\sqrt{1-\frac{|\left<v_{a},\bar{v}_{-a}\right>|^2}{\left|\left<v_{a},v_{a}\right>\right|^2}}}\right]^{\frac{1}{2}}e^{-iArg\left(\left<v_{a},\bar{v}_{-a}\right>\right)},
\end{equation}
with the Sorkin-Johnston modes now given by:
\begin{equation}
u_{a}^{SJ}=\sqrt{\lambda_a}\left(A_{a}v_{a}+B_{a}\bar{v}_{-a}\right).
\end{equation}
The validity of these solutions can be checked by substituting \eqref{eq:Aa1} and \eqref{eq:Ba1} into \eqref{eq:TBOG1}-\eqref{eq:TBOG4}. Notice that this method obviates the need to find $i\Delta$ even though the SJ prescription relies principally upon it in its definition. This will be particularly important for the construction of the SJ state on the Lorentzian cone in Chapter 3.

\section{The Discrete Sorkin-Johnston State}

The continuum framework is now complete. However, it would be remiss not to discuss how the Sorkin-Johnston formalism is applied to the discrete setting since the origination of the concept was motivated thereby. Here we give up the relative generality of the preceding sections and confine our attention to a causal set $\left(C,\prec \right)$ obtained by a $p$ element sprinkling into a $1+1$ dimensional spacetime, $\left(M,g_{\mu \nu} \right)$. The \textit{discrete retarded propagator} has been found in this instance \cite{Johnston:2008za}:

\begin{equation}
R:=\frac{1}{2}C\left( \mathbb{I} + \frac{m^2}{2\rho} C \right)^{-1},
\end{equation}
where $C$ is the causal matrix defined in \eqref{eq:causalmatrix}, $m$ is the mass parameter of the field, $\rho$ is the sprinkling density, and $\mathbb{I}$ is the $p \times p$ identity matrix. The Pauli-Jordan function is found by analogy to the continuum case where $G_A(x,y)=G_R(y,x),$

\begin{equation}
\Delta:=R-R^T,
\end{equation}
and $R^T$ denotes transposition. As before, $i\Delta$ is Hermitian, $(i\Delta)^{\dagger}=(i\Delta)$ and skew-symmetric, $(i\Delta)_{ji}=-(i\Delta)_{ij}$. Hence there is a set of $p$ dimensional eigenvectors now $\{u_a,\bar{u}_a\}$ such that:
\begin{equation} \label{eq:Evector}
(i\Delta)u_a=\lambda_a u_a
\end{equation}
\begin{equation}
(i\Delta)\bar{u}_a=-\lambda_a \bar{u}_a,
\end{equation}
where for example \eqref{eq:Evector} explicitly reads:
\begin{equation}
\sum_j (i\Delta)_{ij}(u_a)_j=\lambda_a (u_a)_i.
\end{equation}
Spectral decomposition proceeds similarly,
\begin{equation}
i\Delta=\sum_a \lambda_a\left(u_a u_a^{\dagger} - \bar{u}_a \bar{u}_a^{\dagger}\right),
\end{equation}
and the Wightman function is given by:
\begin{equation}
W^{SJ}:=Pos[i\Delta]=\sum_a \lambda_a u_a u_a^{\dagger}.
\end{equation}
The discrete Wightman function too will satisfy positivity, the discrete field equation, and orthogonal support.

%%Begin Chapter Three
\chapter{Quantum Field Theory on the Lorentzian Cone}
% \epigraph{She even tried, from what little calculus she'd picked up, to explain it to Franz as $\Delta t$ approaching zero, eternally approaching, the slices of time growing thinner and thinner, a succession of rooms each with walls more silver, transparent, as the pure light of the zero comes nearer...}{Thomas Pynchon, Gravity's Rainbow}

This chapter concerns the explicit construction of the massless scalar Sorkin-Johnston state on the Lorentzian cone. This geometry can be viewed simply as a toy cosmological model. However, similar structures do arise by way of embedding in models of cosmic string inflation \cite{Niedermann:2014yka}. It is therefore important to analyze the effect of such topological defects on extant fields. By implementing a series of coordinate transformations, one is able to able to ``unfurl'' the cone into a flat spacetime, which can alternately be viewed as (i.) a subspace of the Milne universe with the spacelike coordinate periodically identified \cite{9780511622632}, (ii.) an analytic extension of Rindler space into the forward lightcone of the origin, again with the spacelike coordinate periodically identified \cite{Li:1997ka}, or (iii.) a subspace of Misner space restricted solely to within the forward lightcone of the origin, the periodicity of which is inbuilt to its construction \cite{ CBO9780511628863A036}. Mathematically, Misner spacetime's cover is the quotient space of Minkowski spacetime under boost isometries with fixed rapidity parameter \cite{Margalef-Bentabol:2014lta}. The full covering space suffers from pathologies such as chronology horizons and closed timelike curves, and it is generally acknowledged that existence of closed timelike curves leads to unrenormalizable divergences in the stress-energy tensor of quantum fields at the chronology horizons \cite{0264-9381-13-12-002}. These divergences are thought to enforce the chronology protection conjecture by causing the spacetime to be unstable. By requiring that our subspace be within the forward lightcone of the origin, we avoid these pathologies since the closed curves in the spacetime can only be spacelike. It is not obvious then that this region of Misner space should be unstable as the others are. This is what we aim to determine.

%%Probably delete this section
%\section{The Sorkin-Johnston State on Rindler Spacetime}
%The metric on 1+1 Minkowski spacetime is given by
%$$ds^2=(dy^0)^2-(dy^1)^2.$$
%One introduces the Rindler chart via
%$$y^0=a^{-1}e^{a\xi}\sinh(a\eta)$$
%$$y^1=a^{-1}e^{a\xi}\cosh(a\eta),$$
%where $a>0$ is a finite constant, $-\infty<\eta<\infty$, and $-\infty<\xi<\infty.$ This parametrization of Minkowski space restricts the Cartesian chart to $|y^0|<y^1,$ referred to as the right Rindler Wedge (R). One can of course also incorporate the left Rindler Wedge (L), $-|y^0|>y^1$, into the spacetime by including a similar chart with the signs in the above transformation reversed. Since our analysis of the cone will eventually only involve one subregion of the forward wedge of Minkowski space, we will proceed with the SJ state prescription, without loss of generality, only on the right wedge. The metric now takes the form
%$$ds^2=e^{2a\xi}(d\eta^2-d\xi^2),$$
%and the massless scalar field equation becomes
%$$\Box\phi=e^{-2a\xi}\left(\frac{\partial^2}{\partial\eta^2}-\frac{\partial^2}{\partial\xi^2}\right)\phi=0$$

\section{The Sorkin-Johnston State on the Lorentzian Cone}\label{ch1:opts}

We begin with the cosmological metric on the cone
\begin{equation} \label{eq:Milnemetric}
ds^2=dt^2-(at)^2dx^2,
\end{equation}
which describes a $1+1$ dimensional cosmology with $S^0$ creation event at $t=0$ and whose $S^1$ spatial sections grow in time with constant rate, $a$, and have periodicity under the identification $x\sim x+L$. Under the coordinate transformation $t=a^{-1}e^{a\eta}$ with $t\geq0$ (and hence $-\infty<\eta<\infty$), the metric takes on its form in conformal coordinates:
\begin{equation}
ds^2=e^{2a\eta}(d\eta^2-dx^2).
\end{equation}
That this spacetime is diffeomorphic to the restriction of $1+1$ flat Minkowski space to within the forward lightcone of the origin is illuminated by a third transformation:
\begin{equation}
y^0=a^{-1}e^{a\eta}\cosh(ax)
\end{equation}
\begin{equation}
y^1=a^{-1}e^{a\eta}\sinh(ax),
\end{equation}
with $0<y^0<\infty$, $-\infty<y^1<\infty$, and $|y^1|<y^0.$

For the purposes of constructing the SJ state and subsequently finding its stress-energy tensor VEV, we'll use the conformal version of the metric. In these coordinates the massless, minimally coupled $(\xi=0)$ scalar field equation is:
\begin{equation}
\Box\phi=e^{-2ax}\left(\frac{\partial^2}{\partial\eta^2}-\frac{\partial^2}{\partial x^2}\right)\phi=0,
\end{equation}
which being conformally flat,
\begin{equation}
e^{2ax}\Box\phi=\left(\frac{\partial^2}{\partial\eta^2}-\frac{\partial^2}{\partial x^2}\right)\phi=0,
\end{equation}
means that we can select the plane waves,
\begin{equation} \label{eq:MODEZ}
u_{k}(\eta,x)=\frac{1}{\sqrt{2|k|L}}e^{-i(|k|\eta-kx)}
\end{equation}
with periodic boundary condition, $u_{k}(\eta,x+L)=u_{k}(\eta,x)$, enforced by
\begin{equation} \label{eq:BCZ}
k=k_n=\frac{2\pi n}{L},
\end{equation}
as the properly Klein-Gordon normalized, positive frequency mode set with which to construct the SJ state. Henceforth, we will often drop the subscript on $k_n$ under the understanding that it is restricted to the discrete set above. Indeed, in addition to
\begin{equation}
(u_{k},u_{k'})_{KG}=i\delta_{k k'},
\end{equation}
the set $\{u_{k},\bar{u}_k\}$ also satisfies the $L^2(M)$ inner products, defined in \eqref{eq:Lsquaredin}:
\begin{equation} \label{eq:L21}
\left<u_{k},u_{k'}\right>=\left<u_{k},u_{k}\right>\delta_{kk'}
\end{equation}
\begin{equation} \label{eq:L21}
\left<u_{k},\bar{u}_{k'}\right>=\left<u_{k},\bar{u}_{-k}\right>\delta_{k-k'},
\end{equation}
where:
\begin{equation}
\left<u_{k},u_{k}\right>=\frac{1}{2|k|L}\int_{-\frac{L}{2}}^{\frac{L}{2}} dx \int_{-\infty}^{\Lambda} d\eta \, e^{2a\eta}= \frac{e^{2a\Lambda}}{4|k|a}
\end{equation}
\begin{equation}
\left<u_{k},\bar{u}_{-k}\right>=\frac{1}{2|k|L}\int_{-\frac{L}{2}}^{\frac{L}{2}} dx \int_{-\infty}^{\Lambda} d\eta \, e^{2(a+i|k|)\eta}=\frac{e^{2(a+i|k|)\Lambda}}{4|k|(a+i|k|)},
\end{equation}
$\Lambda$ being an integral cutoff in the $\eta$ coordinate, which can be kept finite or, depending upon the reader's proclivity for ice cream, taken to infinity at the conclusion of the calculations we will perform in this section. As mentioned previously, this limiting procedure is not always well defined or unambiguously unique, but for our purposes, it will do. In addition, the eigenvalues are computed using \eqref{eq:EIGENBALUES}:
\begin{equation}
\lambda_k=\frac{e^{2a\Lambda}}{4a\sqrt{a^2+\left|k\right|^2}}.
\end{equation}

These quantities in hand, we can proceed to construct the SJ modes via the Bogolyubov method introduced in Section 2.5:
\begin{equation}
u_{k}^{SJ}=\sqrt{\lambda_k}\left(A_{k}u_{k}+B_{k}\bar{u}_{-k}\right),
\end{equation}
where
\begin{equation}
A_{k}=\frac{1}{\sqrt{2\lambda_k}}\left[1+\frac{1}{\sqrt{1-\frac{|\left<u_{k},\bar{u}_{-k}\right>|^2}{\left|\left<u_{k},u_{k}\right>\right|^2}}}\right]^{\frac{1}{2}}
\end{equation}
and
\begin{equation}
B_{k}=-\frac{1}{\sqrt{2\lambda_k}}\left[-1+\frac{1}{\sqrt{1-\frac{|\left<u_{k},\bar{u}_{-k}\right>|^2}{\left|\left<u_{k},u_{k}\right>\right|^2}}}\right]^{\frac{1}{2}}e^{-iArg\left(\left<u_{k},\bar{u}_{-k}\right>\right)}.
\end{equation}
Noting that $\left<u_{k},u_{k}\right>=\left<u_{-k},u_{-k}\right>$ and using our $L^2(M)$ expressions from above, we find these coefficients to be:
\begin{equation}
A_{k}=\frac{1}{\sqrt{2\lambda_k}}\left[1+\frac{\sqrt{a^2+|k|^2}}{|k|}\right]^{\frac{1}{2}},
\end{equation}
and
\begin{equation}
B_{k}=-\frac{1}{\sqrt{2\lambda_k}}\left[-1+\frac{\sqrt{a^2+|k|^2}}{|k|}\right]^{\frac{1}{2}}\left(\frac{a+i|k|}{\sqrt{a^2+|k|^2}}\right)e^{-2i|k|\Lambda}.
\end{equation}

Now, in order to evaluate the stress energy tensor, we expand our field in terms of these modes:
\begin{equation}
\phi(\eta,x)=\sum_{k=-\infty}^{\infty} (a_{k}u_{k}^{SJ}+a_{k}^{\dagger}\bar{u}_{k}^{SJ}),
\end{equation}
\begin{equation}
a_{k}\left|SJ\right>=0.
\end{equation}
The diagonal terms in the stress-energy tensor are:
\begin{equation}
T_{\eta \eta}=T_{x x}=\frac{1}{2}\left[\left(\frac{\partial\phi}{\partial\eta}\right)^2+\left(\frac{\partial\phi}{\partial x}\right)^2\right],
\end{equation}
while the off diagonal terms are zero. In terms of the SJ modes, the VEV
\begin{equation}
\left<SJ\right|T_{\eta \eta}\left|SJ\right>=\frac{1}{2}\left[\left<SJ\right|\phi,_{\eta}\phi,_{\eta}\left|SJ\right>+\left<SJ\right|\phi,_{x}\phi,_{x}\left|SJ\right>\right],
\end{equation}
reduces to:
\begin{equation}
\left<SJ\right|T_{\eta \eta}\left|SJ\right>=\frac{1}{2} \sum_{k=-\infty}^{\infty} \left[(\partial_{\eta}u_{k}^{SJ})(\partial_{\eta}\bar{u}_{k}^{SJ}) + (\partial_{x}u_{k}^{SJ})(\partial_{x}\bar{u}_{k}^{SJ})\right].
\end{equation}
The intermediate result is fairly simple:
\begin{equation}
\left<SJ\right|T_{\eta \eta}\left|SJ\right>=\frac{\pi \beta}{L^2}+\frac{2\pi}{L^2} \sum_{n=1}^{\infty} \sqrt{\beta^2+n^2},
\end{equation}
where $\beta=\frac{La}{2\pi}$. Notice that the sum is now on the natural numbers $\mathbb{N}$. Also note in particular that our integral regulator $\Lambda$ has dropped out of the calculation. Now, this is clearly a divergent sum as it stands, and moreover, it is not solvable in closed analytic form. However, we can follow a regularization procedure analogous to the one used for an infinite cylinder by introducing a regulator on each term and seeing where it takes us.
\begin{equation}
\left(\left<SJ\right|T_{\eta \eta}\left|SJ\right>-\frac{a}{2L}\right)\rightarrow\frac{2\pi}{L^2} \sum_{n=1}^{\infty} \sqrt{\beta^2+n^2} x^n,
\end{equation}
where $0<x=e^{-\epsilon}<1$, $\epsilon=\frac{\delta2\pi}{L}$, and the regulation parameter, $\delta$, will be taken to zero at the conclusion of the calculation. As $n$ becomes large, terms are exponentially suppressed. This motivates postulating that $\beta$ (or equivalently L) is large, and expanding the square root in a power series about $\beta=\infty$, since the presence of the regulator means any terms where $n$ is nearly as large as $\beta$ will be exponentially suppressed and thus contribute little to the sum. To wit:
\begin{equation}
\sqrt{\beta^2+n^2}=\sum_{m=0}^{\infty} \frac{C_{m}n^{2m}}{\beta^{2m-1}}.
\end{equation}
The $C_m$ are just numerical constants. For example, the first few terms are:
\begin{equation}
=\beta+\frac{n^2}{2\beta}-\frac{n^4}{8\beta^3}+\frac{n^6}{16\beta^5}+...
\end{equation}
In fact, the expansion above converges for any $\beta$ away from the origin. So, as long as the conformal circumference of our cone remains non-zero, we shall feel justified in its validity.

What is now proposed is to calculate $\left<T_{\eta\eta}\right>^{SJ}$ term for term in $m$ and to isolate and subtract the divergent part attributable to the SJ state in the Milne forward wedge, namely $\left<T_{\eta\eta}\right>^{SJ}_{L\rightarrow\infty}$. We will be left with the desired vacuum energy density of the cone:
\begin{equation}
\left<T_{\eta\eta}\right>^{SJ}_{Reg}=\left<T_{\eta\eta}\right>^{SJ}-\left<T_{\eta\eta}\right>^{SJ}_{L\rightarrow\infty}.
\end{equation}

Consider then, the contribution from the $m^{th}$ term in the sum to $\left<T_{\eta\eta}\right>^{SJ}$:
\begin{equation}
\frac{2\pi C_m}{L^2 \beta^{2m-1}} \sum_{n=1}^{\infty} n^{2m} x^n=\frac{(2\pi)^{2m} C_m}{L^{2m+1} a^{2m-1}}\sum_{n=1}^{\infty} n^{2m}x^m
\end{equation}
\begin{equation}
=\frac{(2\pi)^{2m} C_m}{L^{2m+1} a^{2m-1}} Li_{-2m}(x)
\end{equation}
\begin{equation}
=\frac{(2\pi)^{2m} C_m}{L^{2m+1} a^{2m-1}} Li_{-2m}(e^{-\epsilon}),
\end{equation}
where $Li_{-2m}(x)$ is a polylogarithm. We can then expand this about $\epsilon=0$ since we will eventually be taking our regulator to zero \cite{gradshteyn2007}:
\begin{equation}
=\frac{(2\pi)^{2m} C_m}{L^{2m+1} a^{2m-1}} \left(\frac{\Gamma(1+2m)}{\epsilon^{2m+1}}+\sum_{l=0}^{\infty}\frac{\zeta(-(2m+l))}{l!}(-\epsilon)^l\right),
\end{equation}

where $\Gamma(1+2m)$ is the usual gamma function and $\zeta(-(2m+l))$ is the Riemann zeta function. Showing explicitly the first few terms:

\begin{equation}
=\frac{(2\pi)^{2m} C_m}{L^{2m+1} a^{2m-1}} \left(\frac{L^{2m+1}\Gamma(1+2m)}{(\delta 2\pi)^{2m+1}}+\zeta(-2m)-\frac{\zeta(-(2m+1))(\delta 2\pi)}{L}+O\left(\frac{\delta^2}{L^2}\right)\right)
\end{equation}

\begin{equation}
=\frac{(2\pi)^{2m} C_m}{a^{2m-1}} \left(\frac{\Gamma(1+2m)}{(\delta 2\pi)^{2m+1}}+\frac{\zeta(-2m)}{L^{2m+1}}-\frac{\zeta(-(2m+1))(\delta 2\pi)}{L^{2m+2}}+O\left(\frac{\delta^2}{L^{2m+3}}\right)\right).
\end{equation}

Here, we can see that the only divergent term in the limit $\delta \rightarrow 0$ is the first one. Since this evidently corresponds to the Milne vacuum energy, we remove this term. Now we can safely take the limit and we are left with:

\begin{equation}
=\frac{(2\pi)^{2m} C_m}{a^{2m-1}} \left(\frac{\zeta(-2m)}{L^{2m+1}}\right).
\end{equation}

The reader might now be ecstatic upon noticing that the Riemann zeta function vanishes for any negative, even integer. Hence, the only contribution to the renormalized vacuum energy comes from the $m=0$ term, which with $C_0=1$ and $\zeta(0)=-\frac{1}{2}$ takes the neat value of:

\begin{equation} \label{eq:SESJ}
\left<T_{\eta\eta}\right>^{SJ}_{Reg}=\frac{a}{2L}-\frac{a}{2L}=0.
\end{equation}

\section{Another Route To Stress-Energy}

One criticism, which has been leveled against the use of the Sorkin-Johnston state is that it fails to be Hadamard in a general spacetime \cite{Fewster:2013lqa}. A full discussion of the merits or even necessity of using states, which satisfy the Hadamard condition is outside the scope of this dissertation, but a typical counterargument runs as follows. The Hadamard condition requires that a state's symmetric two point function ought to have the same singularity structure as the corresponding two point function in Minkowski space. The postulate is motivated by the equivalence principle in that the coincidence limit of the two points under consideration should render the propagator as it would look in a locally inertial frame. However, considering that nearly all of quantum gravitational research deals with how physicists need to alter the fundamental kinematical structure underlying spacetime away from the continuum that begets its singularities, the thought that the Hadamard condition will ultimately have any deep physical meaning is tenuous.

Nevertheless, in the case that the maneuvers in the preceding section will strike anyone as doubtful, it will behoove us to consider alternate means of obtaining the renormalized stress-energy tensor. Topically, this will require the use of the Hadamard ansatz and its related regularization scheme. As such, we will proceed \textit{independently} of the Sorkin-Johnston prescription and calculate directly with the conformal modes $\{u^{C}_{k_n},\bar{u}^{C}_{k_n}\}$ in equations \eqref{eq:MODEZ}\footnote{The placement of the $C$ superscript is new, but only to emphasize that these are different than the SJ modes.} and \eqref{eq:BCZ} rather than $\{u^{SJ}_{k_n},\bar{u}^{SJ}_{k_n}\}$ in order to calculate, via Hadamard regularization, the renormalized stress-energy tensor in the conformal vacuum. Begin by restating the definition of Hadamard's elementary function in terms of Wightman functions:
 \begin{equation}
 G^{(1)}(X,Y)=W(X,Y)+\bar{W}(X,Y),
 \end{equation}
where $X$ and $Y$ are used to indicate a general coordinate system. As a mode sum in conformal coordinates, the Wightman function reads,

\begin{equation} \label{eq:confwight}
\begin{aligned}
W^C(\eta,x;\eta',x')&=\sum_{n=-\infty}^{\infty} u^C_{k_n}(\eta,x) \bar{u}^C_{k_n}(\eta',x')
\\&=\sum_{n=-\infty}^{\infty} \frac{1}{2|k_n|L}e^{-i|k_n|\eta}e^{ik_nx}e^{i|k_n|\eta'}e^{-ik_nx'}
\\&=\sum_{n=-\infty}^{\infty}\frac{1}{2|k_n|L}e^{-i|k_n|(\eta-\eta')}e^{ik_n(x-x')}
\\&=\frac{1}{2L} \frac{L}{2 \pi}\sum_{n=-\infty}^{\infty} \frac{1}{|n|} e^{-i|k_n| \Delta \eta}e^{ik_n \Delta x}
\\&=\frac{1}{4 \pi} \left( \sum_{n=1}^{\infty} \frac{\alpha^n}{n} + \sum_{n=1}^{\infty} \frac{\beta^n}{n} + \mathcal{N}_0 \right).
\end{aligned}
\end{equation}
In the step moving to the last line, we have split the sum into its negative and positive integer domains and a divergent, but functionally constant $(n=0)$ term, $\mathcal{N}_0$. The sum on the negative integers was then reoriented by accounting for signs in the complex exponentials, and the definitions,

\begin{equation}
\alpha:=e^{-i\frac{2\pi}{L}(\Delta \eta + \Delta x)}
\end{equation}
\begin{equation}
\beta:=e^{-i\frac{2\pi}{L}(\Delta \eta - \Delta x)},
\end{equation}
were made. The sums for the Wightman function using these definitions are not strictly convergent since the complex exponentials do not lie entirely within the radius of convergence for the Mercator expansion of the complex logarithm, namely:

\begin{equation}
-\ln(1-z)=\sum_{n=1}^{\infty} \frac{z^n}{n},
\end{equation}
which is convergent so long as $|z|\leq1$ and $z \neq 1$. In order to get the sums in \eqref{eq:confwight} to converge, insert a regulator $e^{-\epsilon}$ next to both $\alpha$ and $\beta$ in each term and then take $\epsilon \in \mathbb{R}$ to zero at the calculation's conclusion. With this mechanism, both sums converge to the principle branch of the complex logarithm:

\begin{equation}
W^C(\eta,x;\eta',x')= \frac{1}{4 \pi} \left( -\ln(1-\alpha e^{-\epsilon}) -\ln(1-\beta e^{-\epsilon}) + \mathcal{N}_0 \right).
\end{equation}
The Hadamard elementary function is then
\begin{equation}
\begin{aligned}
G^{(1)C}(\eta,x;\eta',x')&=-\frac{1}{4\pi} \left( \ln \left[(1-\alpha e^{-\epsilon})(1-\beta e^{-\epsilon})(1-\bar{\alpha} e^{-\epsilon})(1-\bar{\beta} e^{-\epsilon})\right] -2\mathcal{N}_0 \right)
\\&=-\frac{1}{4\pi} \left( \ln \left[ (2-\Re[\alpha]e^{-\epsilon}-O(\epsilon))(2-\Re[\beta]e^{-\epsilon}-O(\epsilon)) \right] \right).
\end{aligned}
\end{equation}
At this point we will consider it safe to take $\epsilon$ to zero since the argument of the logarithm is entirely real. After a little bit of trigonometric manipulation, the result is:

\begin{equation} \label{eq:cylindhad}
G^{(1)C}(\eta,x;\eta',x')=-\frac{1}{4 \pi} \left( \ln\left[ 16\sin^2(\frac{\pi}{L}(\Delta \eta + \Delta x))\sin^2(\frac{\pi}{L}(\Delta \eta - \Delta x)) \right]  -2 \mathcal{N}_0  \right).
\end{equation}
Note that this calculation has been nearly identical to the treatment given to the infinite cylinder in section 4.2 of \cite{ 9780511622632}. Elements of that treatment will find further use in the remainder of the calculation.

Now, a function is Hadamard if in the limit $X \rightarrow Y$, it can be put into the form of the Hadamard ansatz. In two dimensions, this is \cite{Ghaffarnejad:2014tja}:

\begin{equation} \label{eq:ansatz}
G^{(1)}(X,Y) = v(X,Y)\ln \sigma(X,Y) + w(X,Y),
\end{equation}
where $\sigma(X,Y)$ is one half the square of the geodesic distance between $X$ and $Y$. Here also, $v(X,Y)$ and $w(X,Y)$ are nonsingular, smooth functions. If the Hadamard function is initially unknown, it's coincidence form can ostensibly be found by expanding $v(X,Y)$ and $w(X,Y)$ as:

\begin{equation}
v(X,Y)=\sum_{n=0}^{\infty} v_n(X,Y)\sigma^n
\end{equation}
\begin{equation}
w(X,Y)=\sum_{n=0}^{\infty} w_n(X,Y)\sigma^n,
\end{equation}
operating on the Hadamard ansatz with the Klein-Gordon operator, and solving for the expansion coefficients recursively using the fact that $G^{(1)}$ solves the field equation. Since we have an explicit expression for $G^{(1)C}$ in conformal coordinates, we simply take the limit $(\eta,x) \rightarrow (\eta',x'),$ or equivalently, $\left(\Delta \eta, \Delta x \right) \rightarrow 0.$ We can then expand the sin functions in the logarithm about a zero argument. This gives:

\begin{equation} \label{eq:limHAD}
\begin{aligned}
G^{(1)C}(\eta,x;\eta',x') &\cong -\frac{1}{4 \pi} \left( \ln\left[ 16\left(\frac{\pi}{L}\right)^4(\Delta \eta + \Delta x)^2(\Delta \eta - \Delta x)^2 \right]  -2 \mathcal{N}_0  \right)
\\&=-\frac{1}{4\pi} \left( 2\ln \left[ \frac{\left( \Delta \eta^2 - \Delta x^2 \right)}{L^2} \right] + 4 \ln(2\pi) - 2\mathcal{N}_0 \right).
\end{aligned}
\end{equation}
This is beginning to look familiar. Note that there is an $L^2$ in the denominator of the first logarithm in order to keep the argument dimensionless. For any reasonable yarmulke or cone, $L$ will be nonzero and as such will not ruin the singularity structure. in order to get this into its correct form, add and subtract $\frac{1}{2\pi} \ln[e^{2a\eta'}]$ to \eqref{eq:limHAD}. It will then read:

\begin{equation} \label{eq:HADAbard}
G^{(1)}(\eta,x;\eta',x')=-\frac{1}{4\pi} \left( 2\ln \left[ \frac{ e^{2a\eta'} \left( \Delta \eta^2 - \Delta x^2 \right)}{L^2} \right] + 4 \ln(2\pi) - 2\mathcal{N}_0 - 2 \ln[e^{2a\eta'}] \right).
\end{equation}
Finally, notice that in the true coincidence limit, the geodesic distance squared is just the familiar expression for the line element in our spacetime:

\begin{equation}
\begin{aligned}
\lim_{(\Delta \eta, \Delta x) \rightarrow 0} \sigma^2 (\Delta \eta, \Delta x) &= ds^2 \\&= e^{2a\eta'} (d\eta'^2 - dx'^2)
\\&= e^{2a\eta'} (\Delta \eta^2 - \Delta x^2).
\end{aligned}
\end{equation}
To lowest order in $n$, one can then identify \eqref{eq:HADAbard} with \eqref{eq:ansatz} and the smooth functions as:

\begin{equation}
v_0(\eta,x;\eta',x')=-\frac{1}{\pi}
\end{equation}
\begin{equation}
w_0(\eta,x;\eta',x')=-\frac{1}{4\pi}(4 \ln(2\pi) - \mathcal{N}_0 - 2 \ln[e^{2a\eta'}]).
\end{equation}

Generally, the way Hadamard renormalization then proceeds is by differentiating in the point splitting limit and removing the previously logarithmically singular term in order to get the stress-energy tensor.

\begin{equation}
\left<T_{\mu \nu} \right>_{Reg} =\frac{1}{2} \lim_{X \rightarrow Y} \mathcal{D}_{\mu(X) \nu(Y)} G_{Reg}^{(1)}(X,Y),
\end{equation}
where $G_{Reg}^{(1)}(X,Y)=w(X,Y)$. The bilocal differential operator is given in minimally coupled, massless case by \cite{PhysRevD.56.4633}:

\begin{equation} \label{eq:diffop}
\mathcal{D}_{\mu(\eta,x) \nu(\eta',x')}=\frac{1}{2} \{ g^{\mu'}_{\mu} \nabla_{\mu'} \nabla_{\nu} + g^{\nu'}_{\nu} \nabla_{\mu} \nabla_{\nu'} - g_{\mu \nu} g^{\beta'}_{\beta} \nabla_{\beta'} \nabla^{\beta} \},
\end{equation}
where $g^{\beta'}_{\beta}$ is the bivector of parallel transport, which when $X \rightarrow Y$, limits as $g^{\beta'}_{\beta}\rightarrow \delta^{\beta'}_{\beta}$. With the understanding that $\mathcal{D}$ will be operating on the biscalar Hadamard function, \eqref{eq:diffop} reads:

\begin{equation}
\mathcal{D}_{\mu \nu}=\frac{1}{2} \{ g^{\mu'}_{\mu} \partial_{\mu'} \partial_{\nu} + g^{\nu'}_{\nu} \partial_{\mu} \partial_{\nu'} - g_{\mu \nu} g^{\beta'}_{\beta} \partial_{\beta'} \partial^{\beta} \},
\end{equation}
and the diagonal components are:

\begin{equation}
\mathcal{D}^C_{\eta \eta} = \mathcal{D}^C_{x x} = \frac{1}{2} \{ g^{\eta'}_{\eta} \partial_{\eta'} \partial_{\eta} + g^{x'}_{x} \partial_{x'} \partial_{x} \},
\end{equation}
with the off-diagonal components again zero. Applying this to \eqref{eq:cylindhad} and following \cite{ 9780511622632} in that $\partial_{\mu}(\mathcal{N}_0)=0$, we get:

\begin{equation}
\mathcal{D}^C_{\mu \nu}G^{(1)C}=-\frac{\pi}{2L^2}\left(  \csc^2\left( \frac{\pi}{L}(\Delta \eta + \Delta x) \right) + \csc^2\left( \frac{\pi}{L}(\Delta \eta - \Delta x) \right)  \right)(g^{\eta'}_{\eta} + g^{x'}_{x}).
\end{equation}
Upon expanding both squared cosecant functions,

\begin{equation}
\csc^2\left( \frac{\pi}{L}(\Delta \eta \pm \Delta x) \right)=\frac{L^2}{(\pi(\Delta \eta \pm \Delta x))^2}+\frac{1}{3} + O((\Delta \eta \pm \Delta x)^2),
\end{equation}
we observe that the divergent term in the stress-tensor corresponds to the logarithmic term in the coincident Hadamard function. Once these are removed, we can take the coincidence limit. Terms of all higher orders vanish, and we are left with:

 \begin{equation} \label{eq:SEconf}
 \left<T_{\eta \eta} \right>^C_{Reg} =-\frac{\pi}{6L^2},
 \end{equation}
 which is a slightly different result than that calculated in the previous section\footnote{Note that if we had used the pure conformal modes instead of the SJ modes in the renormalization method of 3.1, the result would have been identically $-\frac{\pi}{6L^2}$. This speaks to the robustness of the different regularization methods on the same vacuum.}. We turn to this presently.
%%look at that paper and see if it is possible to find Hadamard function of our SJ state and put it in Hadamard form.

\chapter{Conclusion}
% \epigraph{A thing is mighty big when time and distance cannot shrink it.}{Zora Neale Hurston, Tell My Horse: Voodoo and Life in Haiti and Jamaica}

\section{Beyond the Zero}

With two contending expressions now available for the vacuum energy of the Lorentzian cone, a more general discussion of what might be expected for conical topology change is in order. The first issue to address is the discrepancy between \eqref{eq:SESJ} and \eqref{eq:SEconf}. While the classical stress-energy tensor is by construction a covariant quantity, the renormalized expectation value of the quantal stress-energy tensor is generally not \cite{finster2012quantum}. Moreover, it is not always the case that different regularization schemes give the same result. This, in fact, was part of the original impetus to introduce Hadamard regularization and Wald's axiomatic renormalization \cite{Wald:1978ce}. This was the formalism most closely followed in section 3.2 and the resulting Casimir energy is concordant with the idea that such a topology induces a non-zero vacuum energy.

On the other hand, although Wald's axioms and their corresponding uniqueness theorem are motivated by physically sensible requirements, such as covariant conservation of $\left< T_{\mu \nu} \right>_{Reg}$ and respect for causality, it is not strictly necessary to take up all of these axioms in every instance. For example, the second axiom typically reads \cite{Wald:1978ce}:

\vspace{3mm}

``\textit{In the case of Minkowski spacetime, $T_{\mu \nu}$ is given by normal ordering, $T_{\mu \nu}=:T_{\mu \nu}:$.}''

\vspace{3mm}

However, this axiom is unsuitable for use with topologies of the form $\mathbb{R} \times S^{(n-1)}$ since Minkowski space does not have this topology. For the case of the temporally infinite cylinder, the substitute prescription is to renormalize against the infinite term when the circumference of the cylinder, $L$, is taken to infinity. This is ostensibly the correct prescription for the cylinder, because it becomes effectively $1+1$ dimensional Minkowski space when the limit is taken. But what then is to be done when the limiting case of a topologically non-trivial spacetime is not Minkowski space, such as for the cone or yarmulke?

Here, the limiting open spacetime is the $1+1$ Milne universe, which itself has ambiguities as regards its vacua and thus its vacuum energy \cite{ 9780511622632}. It has both an adiabatic vacuum, which one can write as a linear superposition of Minkowski plane waves, and in which no particles are detected. The other vacuum is the conformal vacuum wherein the energy density has the functional dependence $\left< T_{\eta \eta} \right>^{Mil}_{Reg} = -\frac{1}{24 \pi t^{2}}$ where $t$ is the time coordinate introduced in \eqref{eq:Milnemetric}. The question as to which vacuum we are limiting to is not straightforward. The optimistic resolution would be to say that the null vacuum energy found in the cone with the Sorkin-Johnston modes (as a linear combination of conformal modes) somehow reflects the adiabatic, particle-less vacuum of Milne, while the Casimir energy found with the conformal modes corresponds to the conformal vacuum in Milne with the singularity at creation being somehow stabilized by the topological structure of the conical cobordism. However, these suggestions are by no means sure.

Perhaps the most striking aspect of these results is the distinct lack of any \textit{geometrically} singular expressions for the stress-energy expectation on the cone. Although one might expect the cone to inherit the creation singularity from the Milne universe, this at first glance does not seem to be the case. Furthermore, there looks to be no pathological energy burst such as is witnessed in the trousers. One however could reason that since the Milne conformal vacuum, where there is no Casimir energy, should be the limiting case of the Misner conformal vacuum, we should add $\left< T_{\eta \eta} \right>^{Mil}_{Reg} = -\frac{1}{24 \pi t^{2}}$ to \eqref{eq:SEconf}. This seems rather ad hoc, but plausible. Even still, this divergence decays to the causal future of the singularity, making it of a distinctly different nature than the trouser singularity. Conceptually, an observer at any point to the causal future of the trouser crotch will be besieged by an infinite bath of radiation, while an observer on the cone will witness either an ever cooling universe\footnote{Of course semiclassically, the $\left<T_{\eta \eta}\right> \propto t^{-2} $case would have a fairly significant back-reaction on the spacetime. Moreover, at times close to $t \sim t_p$, the Planck time, the creation event should actually be described by quantum gravity proper. This analysis assumes some temporal cutoff at around $t_p$ then and that the deleterious effects of a decaying energy burst will not be so severe on the $S^1$ topology as would a $\delta^2$ type divergence.}, one with homogenous negative pressure, or one devoid of any detectable matter at all. As a side note, only the first two instances are consistent with the common appearance of a trace anomaly in the stress-energy VEV of a conformally invariant field theory \cite{Wald:1978ce}.

This brings us back to Sorkin's conjecture. The causal continuity of the cone and yarmulke can be seen by placing an arbitrary point in the unfurled spacetime of either and considering the future and past lightcones emanating therefrom. As the point is imagined to be pulled closer and closer to the singularity, one lightcone diminishes smoothly to zero while the other transitions smoothly to illuminate the entire yarmulke or cone. At no point does the volume of the future or past lightcone change abruptly. In this light\footnote{Pun intended.}, our results that the cone displays no irreparable divergences may be viewed as some measure of corroboration for the Sorkin conjecture. Put differently, there is another conjecture due to Sorkin and Borde that in $n$ dimensions, only Morse metrics on topological cobordisms with index $1$ or $n-1$ critical points are causally discontinuous \cite{Dowker:2002hm}. Since the yarmulke and cone have Morse indices $\lambda=2$ and $\lambda=0$ respectively with only one critical point each, we can rephrase our result. The mild or non-singularity of $\phi$ on the cone is consistent with with the fact that the Morse index of its critical point $t=0$ is neither $1$ nor $n-1=2-1=1$, which correspond to the trousers. This signifies roughly that conical cobordisms at the quantum level are at least stable in terms of intrinsic geometry in the semiclassical limit. It would be interesting in future work to examine whether coupling to the Ricci scalar $(\xi \neq 0)$ or allowing the field to be massive remains consistent with these observations.

One corollary of this analysis is that we can now revisit the question posed at the outset of this dissertation. Notwithstanding the ill-definition of the semiclassical vacuum, we have actually been able to address a notion of nothingness one level deeper. While we initially described the cone as an $S^0 \rightarrow S^1$ cobordism, it can in fact be viewed as an $\{ \, \} \rightarrow S^1$ cobordism\footnote{Note that this does not cohere with the definition of cobordism in terms of manifold topology, but we could make a similar definition using causal sets. In addition, all $S^n$ are cobordant in this sense with $\{ \, \}$.} \cite{Dowker:2002hm}. Mathematically, the empty set is perhaps the closest thing we can define that approaches the concept of nothing. But if $S^1$ is cobordant with it, and the $S^1$ vacuum energy is the negative Casimir energy found in section 3.2, then assuming that the scalar Hamiltonian VEV in $\{ \, \}$ is zero, the existence of such structures is actually \textit{preferred} and nothingness will be unstable. Then, we might rest easy at night knowing that the once dark void is little more than a gently lilting sea of cones and yarmulkes.

%talk about Fay's classification of good/bad cobordisms
%maybe a little bit of morse theory?
%explain lack of causal discontinuity
%bring it back to sorkin's conjecture

%trace anomaly?

%\begin{thebibliography}{123}
%\addcontentsline{toc}{chapter}{Bibliography}
%\raggedright

%\bibliography{MainBibTex}

\bibliographystyle{ieeetr}

\bibliography{icldiss}

%\end{thebibliography}

\end{document}